\documentclass[%
 aip,
 amsmath,amssymb,
reprint,
onecolumn 
]{revtex4-1}

\usepackage{amsmath,amsthm,amsfonts,amssymb,times,bbm,graphicx,mathtools}
\usepackage{xcolor}
\usepackage{hyperref}

\DeclareMathOperator{\End}{End}
\newcommand{\one}{\mathbbm{1}}

\usepackage{tikz}
\usetikzlibrary{shapes.geometric, arrows}

\tikzstyle{box} = [rectangle, 
minimum width=2cm, 
minimum height=1cm,
text centered, 
draw=black]

\tikzstyle{nobox} = [rectangle,
minimum height=1cm,
text centered, 
draw=none]

\tikzstyle{small} = [rectangle,
text centered, 
draw=none]

\tikzstyle{arrow} = [->,double equal sign distance, >=implies, shorten >= 8pt, shorten <= 8pt]
\tikzstyle{darrow} = [<->,double equal sign distance, >=implies, shorten >= 8pt, shorten <= 8pt]

\newtheorem{theorem}{Theorem}
\newtheorem*{theorem*}{Theorem}

\newtheorem{lemma}[theorem]{Lemma}
\newtheorem{corollary}[theorem]{Corollary}
\newtheorem{definition}[theorem]{Definition}
\newtheorem{example}[theorem]{Example}

\newcommand{\bra}[1]{\mbox{$\langle #1 |$}}
\newcommand{\ket}[1]{\mbox{$| #1 \rangle$}}

\newcommand{\trip}[2]{\mbox{$\tr \left( #1 \  #2 \right)$}}

\makeatletter
\newcommand{\ssymbol}[1]{^{\@fnsymbol{3}}}
\makeatother

\newcommand{\mc}[1]{\mathcal{#1}}
\newcommand{\mr}[1]{\mathrm{#1}}
\newcommand{\mb}[1]{\mathbbm{#1}}
\newcommand{\HermH}{\Herm\left( \mc{H}\right)} 

\def\ZZ{\mb{Z}}
\def\RR{\mb{R}}
\def\QQ{\mb{Q}}
\def\CC{\mb{C}}

\def\NN{\mb{N}}

\def\H{\mc{H}}
\newcommand{\Aut}[1]{\mbox{$\mr{Aut}\left(#1\right)$}}
\newcommand{\LinP}[1]{\mbox{$\mr{Lin}\left(#1\right)$}}
\def\Acal{\mc{A}}

\def\Id{\mb{1}}

\newcommand{\muf}{{\mu_1^Q}} 

\DeclareMathOperator{\Wig}{Wig}
\DeclareMathOperator{\Kad}{Kad}
\DeclareMathOperator{\Hil}{Hil}
\DeclareMathOperator{\Jor}{Jor}
\DeclareMathOperator{\HilR}{Hil_\RR}
\DeclareMathOperator{\JorR}{Jor_\RR}
\DeclareMathOperator{\Lin}{Lin}  
\DeclareMathOperator{\Aff}{Aff}

\DeclareMathOperator{\Harm}{Harm}

\DeclareMathOperator{\conv}{conv}
\DeclareMathOperator{\aff}{aff}
\DeclareMathOperator{\dir}{dir}
\DeclareMathOperator{\lin}{span}
\DeclareMathOperator{\tr}{tr}
\DeclareMathOperator{\Tr}{tr} 

\DeclareMathOperator{\Span}{span}

\DeclareMathOperator{\Sym}{Sym}

\newcommand{\stp}[1]{\mbox{$\mr{sp}_{#1}$}}
\newcommand{\STP}[1]{\mbox{$\mr{SP}_{#1}$}}
\newcommand{\ssc}[2]{\pi_{#1}^{#2}}
\newcommand{\SSC}[2]{\Pi_{#1}^{#2}}
\newcommand{\SSA}[2]{\Pi_{{#1}+\bold{#2}}}
\newcommand{\Stab}[1]{\mbox{$\mr{Stab}_{#1}$}}
\newcommand{\Cl}{{\mc{C}_{n,d}}} 
\newcommand{\Cle}{{\mc{C}_{n,d}^{(\rm{G})}}} 

\def\Herm{\mr{Herm}}

\newcommand{\Sp}{\mr{Sp}(\ZZ_d^{2n})} 

\begin{document}

\title{Wigner's Theorem for stabilizer states and quantum designs}

\author{Valentin Obst}
\affiliation{Fraunhofer FKIE, Wachtberg, Germany
}
\affiliation{Institute for Theoretical Physics, University of Cologne, Germany}
\author{Arne Heimendahl, Tanmay Singal, David Gross}
 \affiliation{Institute for Theoretical Physics, University of Cologne, Germany}

\date{\today}

\begin{abstract}
	We describe the symmetry group of the stabilizer polytope for any number $n$ of systems and any prime local dimension $d$.
	In the qubit case, the symmetry group  coincides with the linear and anti-linear Clifford operations. 
	In the case of qudits, the structure is somewhat richer:
	For $n=1$, it is a wreath product of 
	permutations of bases and permutations of the elements within each basis.
	For $n>1$, the symmetries are given by \emph{affine symplectic similitudes}.
	These are the affine maps that preserve the symplectic form of the underlying discrete phase space up to a non-zero multiplier.
	We phrase these results with respect to a number of a priori different notions of ``symmetry'',
	including \emph{Kadison symmetries} (bijections that are compatible with convex combinations),
	\emph{Wigner symmetries} (bijections that preserve inner products),
	and symmetries realized by an action on Hilbert space.
	Going beyond stabilizer states, we extend an observation of Heinrich and Gross \cite{Heinrich2019robustnessofmagic} and show that the symmetries of fairly general sets of Hermitian operators are constrained by certain moments.
	In particular: 
	the symmetries of a set that behaves like a 3-design preserve Jordan products and are therefore realized by conjugation with unitaries or anti-unitaries. 
	(The structure constants of the Jordan algebra are encoded in an order-three tensor,
	which we connect to the third moments of a design).
	This generalizes Kadison's formulation of the classic Wigner Theorem on quantum mechanical symmetries.
\end{abstract}


\maketitle


\section{Introduction}

\subsection{Motivation}

The study of symmetries in quantum mechanics has a long and successful history. 
At the same time, the stabilizer formalism is central to large parts of quantum information theory.
In this paper, we combine these two concepts and study the symmetries of the polytope whose vertices are the projections onto stabilizer states.

We feel that the question has a natural appeal, but also list two more pragmatic applications below.

\subsubsection{Estimating robustness of magic}

It is strongly believed that the worst-case time complexity of simulating a quantum computer operating on $n$ qubits or qudits scales exponentially in $n$.
There are, however, a surprising number of special cases for which non-trivial classical algorithms achieve a significantly better performance.

The best-known case is given by the Gottesman-Knill Theorem, which states that quantum computations that start in a stabilizer state and apply only Clifford operations and Pauli measurements can be simulated in time polynomial in $n$ \cite{Gottesman1998,Aaronson2004} (precise definitions will be given in Sec.~\ref{sec:stab-formalism}).
To extend these results to mixed states, assume that the computer starts in a state $\rho$ that is a convex combination
\begin{align*}
	\rho = p |s_1\rangle\langle s_1| + (1-p) |s_2\rangle\langle s_2|
\end{align*}
of two stabilizer states.
A Monte Carlo simulation algorithm immediately suggests itself:
with probability $p$, output the result obtained by applying the Gottesman-Knill method to the input  $|s_1\rangle$;
with probability $1-p$ do the same for input $|s_2\rangle$.
This procedure will then sample from the same distribution as the quantum computer operating on $\rho$.
This observation motivates the analysis of the convex hull of all stabilizer states: the \emph{stabilizer polytope} (c.f.\ e.g.\ \cite{virmani2005classical,seddon2019quantifying,seddon2021quantifying,heimendahl2022axiomatic}).

The algorithm can be generalized further to deal with input states $\rho$ that lie outside of the stabilizer polytope.
In this case $\rho$ can be still be written as an \emph{affine} combination of stabilizer states, where, however, some of the coefficients may become negative.
It turns out that the runtime of the generalized Monte Carlo algorithms can be expressed as a function of certain distance measures between the state and the stabilizer polytope -- e.g.\ the \emph{robustness of magic} \cite{Campbell2017,Heinrich2019robustnessofmagic}.

Computing the robustness of magic for $n$ qubit input states (even for product states)  amounts to solving a linear program in an exponentially large space -- a highly non-trivial task.
Reference~\cite{Heinrich2019robustnessofmagic} showed that one can achieve a significant speed-up for this problem by exploiting the symmetries of the stabilizer polytope.
This paper included the first systematic analysis of the symmetry group known to us.

\subsubsection{Generalized quasi-probability distributions}

In addition to the stabilizer polytope, other ``polytopes with efficiently simulable vertices'' have been used to design Monte Carlo algorithms.
Indeed, in odd dimensions, there is a discrete version of the Wigner function on which Clifford operations act by permuting points \cite{gross_s_2006}.
A Wigner function analogue of the Gottesman-Knill method says that one can efficiently convert a Clifford circuit into the corresponding permutation and also efficiently describe Pauli measurements
\cite{veitch_negative_2012,mari2012positive,veitch_resource_2014,pashayan2020estimation}. 
Now consider an input state $\rho$ whose discrete Wigner function is a probability distribution on phase space.
In close analogy to the previous algorithm, one can sample a single point according to the distribution, track its path through phase space  under the influence of the Clifford circuits and finally obtain a result for any Pauli measurement -- all in such a way that the expected value equals the expected value of the measurements if they were conducted on a quantum computer.

Unfortunately, a Clifford-covariant Wigner function does not exist for qubits.
As an alternative, it has been proposed to use the vertices of the \emph{polar dual polytope} to the stabilizer polytope as a substitute (in the literature often referred to as the $ \Lambda $-polytope)~\cite{ZurelRaussendorf2020,OkayRaussendorf2021,Heimendahl2019,rau20sim}.
The vertices are operators which, by definition of convex duality, have non-negative inner products with all stabilizer states.
They are also permuted by Clifford operations, so that an analogous simulation algorithm seems achievable.
One issue with implementing this program lies in the fact that the vertices of the polar dual polytope (or, equivalently, the facets of the stabilizer polytope) are not explicitly known except for qudit single systems \cite{gott06class} and two qubits \cite{Reichhardt2005}.
As any polar dual polytope shares the symmetry group with the primal one, 
 it is our hope that the present paper contributes to a better understanding of the $\Lambda$-polytope.


\subsection{Stabilizer Formalism}\label{sec:stab-formalism}

Here, we briefly introduce the central notions of the stabilizer formalism, focusing on its relation to symplectic geometry.
For more background, 
see
Ref.~\cite[Sec.~10.5]{NielsenChuang2011}
and
Refs.~\cite{Gottesman1998, appleby_symmetric_2005, gross_s_2006,gross2021schur}.

The details of the definitions differ somewhat between the case of even- and odd-dimensional Hilbert spaces.
Since all original contributions of this paper pertain to the odd-dimensional case, we only state those definitions in detail.

\subsubsection{The Pauli group}
Our starting point is the \emph{Pauli group}.
For $d$ an odd prime, consider the Hilbert space $\CC^d$ with basis $\{ |q\rangle \,|\, q\in\ZZ_d \}$, where $\ZZ_d$ denotes set of integers modulo $d$.
The construction makes use of a $d$-th root of unity $\omega=\mr{exp}(i2\pi/d)$, and one of its square roots, namely $\tau:=\omega^{(d+1)/2}$.
The generalizations of the familiar Pauli $\sigma_x$ and $\sigma_z$ matrices are defined by their actions on basis vectors as
\begin{equation*}
  X\ket{q}=\ket{q+1},\quad Z\ket{q}=\omega^{q}\ket{q}.
\end{equation*}
Now consider the composition of $n$ systems, with Hilbert space
$(\mb{C}^d)^{\otimes n}$.
For $\bold{a}_X, \bold{a_Z} \in \mb{Z}_d^n$, define the \emph{Weyl-Heisenberg operator}
or \emph{generalized Pauli operator} as
\begin{equation}\label{eq::def_HW_ops}
       T(\bold{a}_X,\bold{a}_Z)
       :=\tau^{-\bold{a}_X^{\mathsf{\top}}\bold{a}_Z}
       \bigotimes_{i=1}^n Z^{(\bold{a}_Z)_i}X^{(\bold{a}_X)_i}.
\end{equation}
Often, the labels $\bold{a}_X$ and $\bold{a}_Z$ are concatenated to a single vector $\bold{a} = (\bold{a}_X, \bold{a}_Z)\in\mb{Z}_d^{2n}$. 
The Weyl-Heisenberg operators generate the \emph{Pauli group}
 \begin{equation*}
  \mc{P}_{n,d}=\left\{\tau^{b}\,T(\bold{a})\,\big\vert\,b\in\ZZ_d\text{ and }\bold{a}\in\ZZ_d^{2n}\right\}.
\end{equation*}

The Weyl-Heisenberg operators obey the composition law
\begin{align}
	\begin{split}
		T(\bold{a})T(\bold{b})&=\tau^{-[\bold{a},\bold{b}]}T(\bold{a}+\bold{b}), \label{eqn:pauli comp}
  \end{split}
\end{align}
and the commutation law
\begin{align}
	\begin{split}
		T(\bold{a})T(\bold{b})&=\omega^{-[\bold{a},\bold{b}]}T(\bold{b})T(\bold{a}). 
  \end{split}
\end{align}
Both laws involve the \emph{symplectic form} $ [\cdot, \cdot]: \ZZ_d^{2n} \times \ZZ_d^{2n} \to \ZZ_d $ defined by
\begin{equation*}
   [\bold{a},\bold{b}]:=\bold{a}_X^{\mathsf{\top}}\bold{b}_Z-\bold{b}_X^{\mathsf{\top}}\bold{a}_Z.
\end{equation*}
Here, we see the first instance of the important relationship between the stabilizer
formalism and symplectic geometry.
The finite vector space $\ZZ_{d}^{2n}$, equipped with the form $[\cdot, \cdot]$ will be referred to as the \emph{phase space} associated with the Hilbert space $(\CC^d)^{\otimes n}$.

The Weyl-Heisenberg operators form a basis of the vector space $L\big((\CC^d)^{\otimes n}\big)$ of linear operators on the Hilbert space.
What is more, this space is a Hilbert space in its own right, when equipped with the \emph{Hilbert-Schmidt} inner product
\begin{align*}
 \tr A^\dagger B,
\end{align*} 
Using Eq.~(\ref{eqn:pauli comp}), one may easily verify that the Weyl-Heisenberg operators are orthonormal with respect to this inner product:
\begin{align}\label{eqn:wh basis}
	\tr T(\bold{a})^\dag T(\bold{b}) =d^n \delta_{\bold{a}, \bold{b}}.
\end{align}

\subsubsection{The Clifford group}
\label{sec:clifford}
Next, we introduce the \emph{Clifford group}. 
It can be defined as the normalizer of the Pauli group within the unitary group
\begin{equation*}
  \mc{C}_{n,d}=
	\left\{U\in L\big((\CC^d)^{\otimes n}\big)\,\big\vert\,U^\dagger=U^{-1}\text{ and }U\mc{P}_{n,d}U^\dagger=\mc{P}_{n,d}\right\}
\end{equation*}
Its structure is closely related to the group 
\begin{align*}
	\mr{Sp}(\ZZ_d^{2n}) 
	=
	\left\{ S \in \ZZ_d^{2n\times 2n}\,\big\vert\, [S \bold{a}, S\bold{b}] = [\bold{a}, \bold{b}] \> \forall \bold{a},\bold{b}\in\ZZ_2^{2n} \right\}
\end{align*}
of \emph{symplectic matrices} on the phase space $\ZZ_d^{2n}$.
More precisely, one can show \cite{gross_s_2006} that for any Clifford unitary $U\in\mc{C}_{n,d}$, there exists a symplectic matrix $S \in \mr{Sp}(\ZZ_d^{2n})$ and a vector $\bold{a}\in\ZZ_d^{2n}$ such that 
\begin{equation}\label{eq::CL_on_HW}
  UT(\bold{b})U^\dagger=\omega^{[\bold{a},\bold{b}]}T(S\bold{b}),\quad\forall\;\bold{b}\in\mb{Z}_d^{2n}.
\end{equation} 
Conversely, for any pair of $S\in\mr{Sp}(\ZZ_d^{2n})$ and $\bold{a}\in\ZZ_d^{2n}$ there exists a $U\in\mc{C}_{n,d}$ such that Eq.~\eqref{eq::CL_on_HW} holds.
When $d\ge 3$, $\Cl$ is isomorphic to the semidirect product\cite{Gerardin1977,Neuhauser2002}: $\ZZ_d^{2n} \rtimes \Sp$. Thus any element in $\Cl$ may be written as in the following form: $\omega^\mu T(
\mathbf{a})U_S$, where $\omega \in \ZZ_d$, $\mathbf{a} \in \ZZ_d^{2n}$, and $U_S$ corresponds to the pair $S \in \Sp$ and $\mathbf{a}=\mathbf{0}$ in Eq.~\eqref{eq::CL_on_HW}. Note that the group generated by $U_S$ is isomorphic to $\Sp$.
We will refer to this as the purely symplectic part of the Clifford group. 

This action becomes more transparent when we express it with respect to a different ortho-normal basis of the space $L\big((\CC^d)^{\otimes n}\big)$ of operators.
Indeed, the \emph{phase space point operator basis} $\{ A(\bold{a}) \,|\, \bold{a}\in \ZZ_d^{2n}\}$ arises from the Heisenberg-Weyl basis via a Fourier transform:
\begin{equation}\label{eq::def_PSP}
  A(\bold{a}):=d^{-n}\sum\limits_{\bold{b}\in\mb{Z}_d^{2n}}\omega^{\left[\bold{a},\bold{b}\right]}T(\bold{b}).
\end{equation}
One can easily verify that the Fourier transform turns the linear phase factors in Eq.~(\ref{eq::CL_on_HW}) into translations in the sense that
\begin{equation}\label{eq::CL_on_PSP}
  UA(\bold{b})U^\dagger=A(S\bold{b}+\bold{a}),\quad\forall\;\bold{b}\in\mb{Z}_d^{2n}.
\end{equation} Thus, each Clifford unitary is associated with an affine transformation on phase space whose linear part is symplectic.
To formalize this observation, one defines the \emph{affine symplectic group}
\begin{equation}\label{eq::aff-symp}
  \mr{ASp}(\ZZ_d^{2n}) = \left\{ f: \ZZ_d^{2n} \to \ZZ_d^{2n}, \, f(\bold{b}) = S\bold b+\bold a\,\big\vert\, S \in \mr{Sp}(\ZZ_d^{2n}), \, \bold{a} \in \ZZ_d^{2n} \right\},
\end{equation}
and says that the Clifford group up to phase factors, i.e., $\mc{C}_{n,d}/U(1)$, is isomorphic to $\mr{ASp}(\ZZ_d^{2n})$. 
To state our main Theorem we need the group of all linear maps that preserve the symplectic form up to a non-zero \textit{multiplier}~$\alpha$
\begin{align} \label{eq:symplectic_similitudes}
  \mr{GSp}(\ZZ_d^{2n}) 
	= 
	\left\{ R \in \ZZ_d^{2n\times 2n}\,\big|\, \exists \, \alpha \in \ZZ_d^\times:[R \bold{a}, R\bold{b}] = \alpha [\bold{a}, \bold{b}] \> \forall \bold{a},\bold{b}\in\ZZ_d^{2n} \right\}.
\end{align}
Such maps are sometimes called \emph{symplectic similitudes} \cite{o1978symplectic}.
Adding linear translations in analogy to Eq.~(\ref{eq::aff-symp}), we arrive at the group $\mr{AGSp}(\ZZ_d^{2n})$ of \emph{affine symplectic similitudes}.

Some similitudes have been discussed in the context of the Clifford group before \cite{appleby_symmetric_2005}.
Indeed, the so-called \emph{extended Clifford group} is the group generated by the Clifford unitaries and the anti-unitary map $C$ that performs a complex conjugation in the standard basis of $(\CC^d)^{\otimes n}$.
One immediately verifies that
\begin{align}
\label{eq:complexconjugation}
	C T(\bold a_X, \bold a_Z) C^{-1} = T(\bold a_X, -\bold a_Z).
\end{align}
In other words, complex conjugation corresponds to the phase space map $K_{-1}$ where we define
\begin{align}\label{eqn:Kalpha}
	K_{\alpha} :=
	\begin{pmatrix}
		\Id_n & 0_n \\
		0_n & \alpha\Id_n
	\end{pmatrix},
\end{align}
which is a symplectic similitude for any non-zero $\alpha$.
In fact, it is a straightforward exercise to show that any symplectic similitude can be written as a product of a symplectic matrix and some $K_\alpha$:
\begin{equation*}
	\begin{aligned}
		\mr{GSp}(\ZZ_d^{2n})
	=
	\left\{ SK_\alpha\,\big\vert\, S \in \mr{Sp}(\ZZ_d^{2n})\text{ and } \alpha \in \ZZ_d^\times\right\},
	\end{aligned}
\end{equation*} 
which means that the extended Clifford group is isomorphic to the subgroup of $\mr{AGSp}(\ZZ_d^{2n})$ with multiplier $\pm 1$.

\emph{Remarks}:
\begin{itemize}
	\item
	The expansion coefficients $W_K(a) = \tr A(a)^\dagger K$ of an operator $K$ with respect to the phase space point operators are known as the (discrete) \emph{Wigner function} of $K$ \cite{gross_s_2006}.
  \item
  The connection between the Clifford group and the symplectic group is somewhat more subtle in the qubit case than for odd dimensions.
\end{itemize}

\subsubsection{Stabilizer states}
\label{sec:stabilizer states}

The last fundamental component of the stabilizer formalism are the \emph{stabilizer states}, which we take to be projections onto the simultaneousrt eigenvectors of maximal Abelian subgroups of the Pauli group. 

More precisely, let $L\subset \ZZ_d^{2n}$ be a subset of phase space.
Using Eq.~(\ref{eqn:pauli comp}) one easily sees that the set of Weyl-Heisenberg operators $T(L):=\{ T(\bold b) \,|\, \bold{b}\in L \}$ forms a group up to phase factors if and only if $L$ is a subspace, and that the group is Abelian if and only if the symplectic form $ [\cdot, \cdot] $ vanishes on $L$.
Such subspaces are called \emph{isotropic}.
Isotropic spaces of dimension $n$ -- the largest possible -- are called \emph{Lagrangian}.
It turns out that for a Lagrangian $L$, the group $T(L)$ has a unique joint eigenbasis.
The projection operators onto the elements of this basis arise explicitly in the following way:
choose a linear functional $g: L\to \ZZ_d$.
Then (using Eqs.~(\ref{eqn:pauli comp}, \ref{eqn:wh basis})) one may verify that
\begin{equation}\label{eqn:wh stab projection}
 \SSC{L}{g} \coloneqq \frac{1}{d^n}\sum\limits_{\bold{b}\in L}\omega^{g(\bold{b})}T(\bold{b})
\end{equation}
is a rank-one projection operator onto a joint eigenstate of the operators in $T(L)$.
As $g$ ranges over the set $L^*$ of all $d^n$ linear functionals on $L$, one obtains projections onto an ortho-normal basis.
These are the \emph{stabilizer states}:
\begin{align*}
  \mr{Stab}_{n,d}
	=
	\left\{ \Pi_L^g \,\big|\, L\subset \ZZ_d^{2n}, L\ \text{is a} \text{ Lagrangian and }\; g \in L^* \right\}.
\end{align*}
Using the symplectic form, one can represent any functional $g\in L^*$ explicitly as
\begin{align}\label{eqn:a functional}
	g(\bold{b}) = [\bold{a},\bold{b}], \qquad \forall\, \bold{b} \in L
\end{align}
for some $\bold{a}\in\ZZ_d^{2n}$.
Two different vectors $\bold a, \bold a'$ represent the same functional $g$ if and only if $\bold a - \bold a' \in L$.
This representation is well-suited for transforming the stabilizer state from the Weyl-Heisenberg basis as in Eq.~\eqref{eqn:wh stab projection} into the phase space point operator basis.
A direct calculation shows that
\begin{equation}\label{eq::SS_aff}
  \SSC{L}{g}=
	\frac{1}{d^n}\sum\limits_{\bold{b}\in L+\bold{a}}A(\bold{b}),
\end{equation}
i.e., the expansion coefficients are proportional to the indicator function on the affine space $L+\bold{a}$ (which does not depend on the choice of $\bold{a}$ in Eq.~(\ref{eqn:a functional})).
We can thus equivalently label stabilizer states $\SSC{L}{g}=\SSA{L}{a}$ by affine spaces $L+\bold{a}$ with Lagrangian directional vector space $L$.
With the help of Eq.~(\ref{eq::CL_on_PSP}), one can now easily see that the Clifford group permutes stabilizer states.
Explicitly, if $U$ represents the affine symplectic map $S \cdot + \bold a$, then
\begin{align*}
	U\SSA{L}{b}U^\dagger=\Pi_{S(L+\bold b)+\bold a}.
\end{align*}

The \emph{stabilizer polytope} is defined as the convex hull 
\begin{equation*}
  \STP{d,n}:=\mr{conv}\bigl(\mr{Stab}_{n,d}\bigr)
\end{equation*}
of all stabilizer states.

\subsubsection{The real Clifford group and rebit stabilizer states}\label{subsec:real-stab-states}

An important subclass of Clifford unitaries and stabilizer states are those that only have real-valued entries with respect to some fixed basis, e.g. the standard basis. The \emph{real} Clifford group acting on $ n $-qubits is generated by~\cite{Hashagen2018realrandomized}
\begin{align*}
	\mathcal{C}_{2,n,\RR} := \langle Z_i, H_i, CZ_{ij} \rangle ,
\end{align*}
where the subscripts denote the qubit that the matrix is acting on and $ Z $ is the Pauli $ Z $-matrix, $ H $ the Hadarmard matrix and $ CZ $ the controlled $ Z $ matrix, that is 
\begin{align*}
	Z = \begin{pmatrix}
		1 & 0 \\ 0 & -1
	\end{pmatrix}, \quad H = \frac{1}{\sqrt{2}}\begin{pmatrix}
	1 & 1 \\ 1 & -1 
\end{pmatrix}, \quad CZ = \begin{pmatrix}
1 & 0 & 0 & 0 \\ 0 & 1 & 0 & 0 \\  0 & 0 & 1 & 0 \\ 0 & 0 & 0 & -1
\end{pmatrix}.
\end{align*}
Equivalently, $ \mathcal{C}_{2,n,\RR} $ can be defined as the normalizer of the group generated by the $ n $-fold tensor products of the real Pauli matrices $ X $ and $ Z $~\cite[Sec.~2]{Nebe2001}. 
The real Clifford group generates an orbit 
\begin{align*}
	\{  C\ket{0} \, : \, C \in \mathcal{C}_{2,n,\RR}  \}.
\end{align*}
These vectors are all stabilizer states by definition and we call them the set of \emph{real stabilizer states}.

\subsection{Symmetries}
\label{sec:intro_symmetries}

For the quantum information application that motivated this paper, it is most natural to consider the group of superoperators that preserves the stabilizer polytope.
However, there are many other natural notions of ``symmetry'' one can associate with a set of quantum states.
For example, the discussion of symmetries in quantum physics is traditionally phrased in terms of \emph{Wigner symmetries} -- maps (a priori not necessarily linear) that send pure states to pure states and preserve transition probabilities:
\begin{align*}
	|\psi\rangle\langle\psi| \mapsto |\psi'\rangle\langle\psi'|,
	\qquad
	|\langle\psi'|\phi'\rangle|^2 = |\langle\psi|\phi\rangle|^2 \qquad \forall\, |\phi\rangle, |\psi\rangle \in \H.
\end{align*}

Since the 1960s, a significant body of literature has been created that examines the various notions of ``symmetry of a quantum system'' and their relations.
A monograph is \cite{cassinelli2004theory};
a general treament of the mathematical structure of quantum mechanics with a careful discussion of symmetries is \cite{landsman2017} (in particular: Chapter~5);
for some elementary proofs of Wigner's theorem see \cite{Bargmann, Wick, Simon2008};
the original literature goes back to~\cite{Wigner1931}. 
A common theme is that all natural symmetry groups one can associate with quantum systems tend to be isomorphic, but the isomorphisms are often fairly non-trivial to construct and analyze.

In this paper, we will reprove some of the relationships between different notions of symmetry.
On the one hand, our results are more general than the existing literature, because we treat not only the set of projections $\{|\psi\rangle\langle\psi| \, |\psi\in\mathcal{H}\}$ onto Hilbert space vectors, but symmetries of rather general subsets $Q$ of Hermitian operators on $\mathcal{H}$.
On the other hand, our results are less general, as we restrict attention to finite-dimensional Hilbert spaces.

To guide the intuition, we provide a list of example of sets $Q$ whose symmetry groups can be analyzed by the theory presented below.
(Though we will not perform an analysis for all of them in this paper).

\begin{example}\label{ex:Q}
	Examples of natural sets $Q\subset\Herm(\H)$.
	\begin{enumerate}
		\item
			All pure states on a Hilbert space.
		\item
			All stabilizer states on $n$ qubits.
		\item
			All stabilizer states on $n$ qudits, where $d$ is an odd prime.
		\item
			All stabilizer states on $n$ qudits, where $d$ is composite. 
		\item 
			A set of pure states that form $ d+1 $ mutually unbiased bases (MUBs).
		\item
			A set of pure states that forms a symmetric informationally-complete (SIC) POVM. 
		\item
			All pure states on a Hilbert space that have real-valued coefficients with respect to some fixed basis.
		\item
			All stabilizer states that have real coefficients in the standard basis, as defined in Sec.\ref{subsec:real-stab-states}.
		\item
			The set of phase space point operators $ \{A(\bold{a}), \bold{a} \in  \ZZ_d^{2n} \} $, as defined in \eqref{eq::def_PSP}. 
	\end{enumerate}
\end{example}

\subsubsection{Types of symmetries}

Fix a finite-dimensional Hilbert space $\H$.
Let $L(\H)$ be the set of linear operators on $\H$.
It is a Hilbert space in its own right, when endowed with the Hilbert-Schmidt inner product
\begin{align*}
	A, B \mapsto \tr A^\dagger B.
\end{align*}
Let $\Herm(\H)$ be the real vector space of Hermitian operators. 
If a real structure has been chosen on $\mathcal{H}$, we write $ \Sym(\mathcal{H}) \subset \Herm(\H)$ for the vector space of symmetric operators. 
The Hilbert-Schmidt inner product restricts to a Euclidean inner product on $\Herm(\H)$ and on $\Sym(\H)$.

Let $Q\subset\Herm(\H)$.
With $Q$, associate the sets
\begin{align*}
	\lin(Q)&& \text{its real linear span}, \\
	\aff(Q)&\subset\lin(Q)& \text{its affine hull}, \\
	\conv(Q)&\subset\aff(Q)& \text{its convex hull}, \\
	\dir(Q) && \text{the directional vector space of } \aff{Q}.
\end{align*}

\textbf{Example \ref{ex:Q} continued.} 
In all cases of Ex.~\ref{ex:Q}, except for the two real-valued examples 7.\ and 8., it holds that
$\Span(Q) = \Herm(\H)$ is the vector space of all Hermitian matrices,
$\aff(Q)$ is the affine space of Hermitian matrices of trace 1,
and
$\dir(Q)$ is the vector space of traceless Hermitian matrices.
In the two real-valued cases 7.\ and 8., the above statement holds if ``Hermitian matrix'' is replaced by ``symmetric Hermitian matrix'', i.e. $ \Span(Q) = \Sym(\mathcal{H}) $.

Recall that on $\Herm(\H)$, the \emph{Jordan product} is
\begin{align}\label{eq:jordan_product}
	A\circ B := \frac12(AB + BA),
\end{align}
which preserves the space of Hermitian matrices and the space of symmetric matrices.

We can now introduce the symmetry groups we are interested in.
\begin{definition}\label{def::symmetries}
	Let $Q\subset\Herm(\H)$.
	\begin{enumerate}
		\item
			A bijection $K$ of $Q$ is a \emph{Kadison symmetry} if it can be extended 
			to $\conv(Q)$, in such a way that it  commutes with convex combinations
			\begin{align*} 
				K \left( \sum_{i=1}^k p_i q_i \right) 
				\ = \ 
				\sum_{i=1}^k p_i  \ K(q_i), \ \mathrm{where} \  q_i \in Q, \ p_i \in [0,1], \  \mathrm{and} \ \sum_{i=1}^k p_i=1, 
			\end{align*}
			Let $\Kad(Q)$ be the group of Kadison symmetries.

		\item
			A bijection $\Lambda$ of $Q$ is an \emph{affine symmetry} if it can be extended 
			to $\aff(Q)$, in such a way that it  commutes with affine combinations
  		\begin{align*} 
				 	\Lambda \left( \ \sum_{i=1}^k \ \alpha_i  q_i \ \right)  
					\ = \ 
					\sum_{i=1}^k \alpha_i \Lambda\left( q_i \right), \ 
					\mathrm{where} \ q_i \in Q, \ \alpha_i \in \RR, \ \mathrm{and} \  \sum_{i=1}^k \alpha_i = 1, 
			\end{align*}
			Let $\Aff(Q)$ be the group of affine symmetries.

		\item
			A bijection $L$ of $Q$ is a \emph{linear symmetry} if it can be extended 
			to $\Span(Q)$, in such a way that it  commutes with linear combinations
  		\begin{align*} 
				 	L\left( \ \sum_{i=1}^k \ \alpha_i  q_i \ \right)  
					\ = \ 
					\sum_{i=1}^k \alpha_i L\left( q_i \right), \ 
					\mathrm{where} \ q_i \in Q, \ \alpha_i \in \RR.
			\end{align*}
			Let $\Lin(Q)$ be the group of linear symmetries.

		\item A bijection $W$ on $Q$ that preserves the Hilbert-Schmidt inner product
				\begin{align*}
					\tr\big( W(q_1) W(q_2)\big)
					=
					\tr\big(q_1  q_2\big) 
					\qquad
					\forall\,q_1, q_2\in Q
				\end{align*}
			is a \emph{Wigner symmetry} of $Q$.
			Let $\Wig(Q)$ be the group of Wigner symmetries.

		\item
			A bijection $H$ of $Q$ is a \emph{linear Hilbert space symmetry} 
			if it is of the form $H: q \mapsto Uq U^{-1}$ for some unitary $U$ on $\H$.
			A bijection $H$ of $Q$ is an \emph{anti-linear Hilbert space symmetry} 
			if it is of the form $H: q \mapsto Uq U^{-1}$ for some anti-linear unitary $U$ on $\H$.
			The group generated by these two sets is the \emph{Hilbert space symmetry group} $\Hil(Q)$ of $Q$.
			
		\item
			A  bijection $J$ of $Q$ is a \emph{Jordan symmetry} if it can be extended to a linear isomorphism of  $\Herm(\H)$ that preserves the Jordan product (see Eq.~\eqref{eq:jordan_product}), i.e.,
			\begin{align}\label{eqn:jordan def}
				J(A) \circ J(B) = J(A \circ B)\qquad \forall A, B \in \Herm(\H).
			\end{align}
			Let $\Jor(Q)$ be the group of Jordan symmetries.

		\item
			Assume a real structure has been chosen on $\H$.
			The subgroup of $\Hil(Q)$ that can be implemented with $U$ an orthogonal map is the \emph{real Hilbert space symmetry group} $\HilR(Q)$.
			The subgroup of $\Jor(Q)$ that preserves $\Sym(\H)$ are the \emph{real Jordan symmetries} $\JorR(Q)$.
	\end{enumerate}
\end{definition}


\emph{Remark:} 
All symmetries in Def.~\ref{def::symmetries}, with the exception of the Jordan ones, are defined in terms of actions on $Q$ or on suitable extension (convex, affine, linear) of $Q$. 
In contrast, we require that Jordan symmetries be defined on the entire space of Hermitian matrices.
It might be possible to develop a more minimal approach, where $J$ has to preserve $\circ$ only on the Jordan algebra generated by $Q$.
We did not investigate this weaker requirement here, because it does not satisfy the assumptions used in the existing characterizations of Jordan symmetries that we will build on, and because it would not have made a difference to any of the cases in Example~\ref{ex:Q}.

\section{Results}

\subsection{The symmetries of the stabilizer polytope}

Our main result is a classification of the Kadison symmetry groups of the stabilizer polytopes $\STP{d,n}$. 

There are two types of obvious symmetries of the stabilizer polytope.
First, if $U$ is an element of the Clifford group, then
\begin{align*}
	\Phi_U: \rho \mapsto U \rho U^\dagger
\end{align*}
defines a superoperator which, as discussed in Sec.~\ref{sec:stabilizer states}, permutes stabilizer states and therefore maps $\STP{d,n}$ to itself.
The second type comes from the complex conjugation $C$ in some fixed basis.
While it is \emph{anti-linear} on the Hilbert space, its action on Hermitian operators is given by the transposition,
\begin{align}\label{eqn:conjugate linearization}
	\Phi_{K_{-1}}: \rho \mapsto C\rho C^{-1} = \rho^{\sf T},
\end{align}
which is a \emph{linear} map. 
Because complex conjugation, too, permutes stabilizer states, it follows that $\Phi_{K_{-1}}$ is another symmetry of $\STP{d,n}$. We thus find that the entire extended Clifford group implements symmetries of the stabilizer polytope.

In the appendix of Ref.~\cite{Heinrich2019robustnessofmagic}, it was shown that for qubits, these are \emph{all} the symmetries of $\STP{n,2}$.
What is more, this paper presented a simple example showing that at least for $n=1, d\geq 5$, the symmetry group must be strictly larger. In fact, one can generalize their example by noting that for any symplectic similitude $R$ and vector $\bold{a} \in \ZZ_{d}^{n}$ the linear map 
\begin{equation}
	\begin{aligned}\label{eq::affSim_on_PSP1}
		A(\bold{b})\mapsto A(R\bold{b}+\bold{a}), \quad \forall\, \bold{b}\in \ZZ_d^{2n}
	\end{aligned}
\end{equation}
permutes stabilizer states $\SSA{L}{b}\mapsto\Pi_{R(L+\bold b)+\bold{a}}$. Then, their conclusion follows since for $d\geq 5$ there are similitudes with multiplier different from $\pm 1$.
A full characterization of the remaining cases was left as an open problem.
Here, we settle this issue by arriving at the following classification of the Kadison symmetries for any prime $d$ and any $n$.

\begin{theorem}[Stabilizer Wigner Theorem]\label{thm:main-thm}
	The Kadison symmetry groups of the stabilizer polytopes $\STP{d,n}$ are as follows:
	\begin{enumerate}
		\item \label{item:n=1}
		If $n=1$, the set of stabilizer states can be decomposed as the disjoint union of $d+1$ mutually unbiased bases.
		The symmetry group is the wreath product $S_{d} \wr S_{d+1}$, where $S_{d+1}$ permutes the bases and $S_d^{\times (d+1)}$ permutes the elements within each basis. 
		\item \label{item:complex_jordan}
			If $d=2$ and for $d=3, n > 1$, the symmetries are exactly those that arise by the action by conjugation of the extended Clifford group.
			
		\item \label{item:n>1,d-odd}
			If $d$ is an odd prime and $n>1$, the symmetry group is $\mr{AGSp}(\ZZ_d^{2n})$ acting as in Eq.~\eqref{eq::affSim_on_PSP1} by permuting the phase space point operator basis. 

		\item \label{item:,d=2-rebits}If $ d = 2 $, the symmetry group of the set of \emph{real} stabilizer states is given by the action by conjugation of the real Clifford group.
	\end{enumerate}
\end{theorem}

In Sec.~\ref{sec:proof-main-thm}, we will establish the cases that have not yet been treated in \cite{Heinrich2019robustnessofmagic}, i.e.\ prove the first, third and fourth point of the classification.

The convex hull of the projectors of a set of $ d+1 $ mutually unbiased bases is referred to as the \emph{complementarity polytope} in the literature \cite{Bengtsson2006, Bengtsson2004, Appleby2015GaloisUM, dang2015studies}. 
The fact that the symmetry group of the complementarity polytope has the structure (as a set) of $S_{d+1} \times S_d \times \dots \times S_d$ has been stated in the papers just cited.
The group structure $ S_{d} \wr S_{d+1} $ has been identified before as the combinatorial symmetry group of $ d+1 $ mutually unbiased bases \cite{Gribling2021}.
Note that for the $d=2$, $n=1$ case, the symmetry group, the wreath product structure $S_2 \wr \S_3$ is the same as the extended Clifford group (see Appendix \ref{app:B}).

\emph{Remark:}
It is not difficult to generalize the Stabilizer Wigner Theorem to the situation where the finite fields $\ZZ_d$ for prime $d$ are replaced by finite fields of cardinality the power of a prime.

\subsection{Equivalences between types of symmetries}

We have phrased Thm.~\ref{thm:main-thm} in terms of Kadison symmetries, but in fact, it applies to various other types of symmetries too.

As indicated in Sec.~\ref{sec:intro_symmetries}, 
we will prove relations between different types of symmetries in a way that not only applies to stabilizer states, but to subsets $Q\subset\Herm(\H)$ more generally.
Some relations hold unconditionally.
Others turn out to depend on certain second and third \emph{moments} associated with $Q$.

Starting point of this approach is the following observation, taken from an appendix of Ref.~\cite{Heinrich2019robustnessofmagic}:

\begin{theorem}{}\cite{Heinrich2019robustnessofmagic}
 \label{thm:design_syms}
  Let 
	$\{\psi_i\}_{i=1}^N \subset \CC^d$ 
	be a set of unit vectors.
 Let $L \in \End
 \left(\HermH\right)$
 be a linear map on Hermitian operators that permutes the projectors $\{\ket{\psi_i}\bra{\psi_i}\}_{i=1}^N$. 
 \begin{enumerate}
 	\item
		If $\{\psi_i\}_{i=1}^N$ is a complex projective 1-design, then $L$ is unital (i.e.\ $L(\one)=\one$).
 	\item
	  If $\{\psi_i\}_{i=1}^N$ is a complex projective 2-design, then $L$ is orthogonal and trace-preserving.
 	\item
	  If $\{\psi_i\}_{i=1}^N$ is a complex projective 3-design, then $L$ is of the form $L= U\cdot U^\dagger$, where $U$ is either a unitary or an antiunitary. 
\end{enumerate}
\end{theorem}

As stated, this prior result is not applicable e.g.\ to real stabilizer states. 
In Section~\ref{sec:symmetry-proofs}, we establish a collection of much more general result, which may be of independent interest.

\begin{theorem}[Relation between symmetries, informal summary]\label{thm:symmetry_summary}
	For $Q\subset\Herm(\H)$, we have the following relations between the various symmetry groups:
\begin{center}
	\begin{tikzpicture}[node distance=4cm]
		\node (node1) [nobox] {$\Kad(Q)\Longleftrightarrow \Aff(Q)$};
		\node (node2) [nobox, right of=node1] {$\Lin(Q)$};
		\node (node3) [nobox, right of=node2] {$\Wig(Q)$};
		\node (node4) [nobox, right of=node3] {$\Jor(Q)\Longleftrightarrow\Hil(Q)$.};

	\draw[arrow] ([yshift=2mm]node1.east) -- node[anchor=south] {$0\neq\aff(Q)$} ([yshift=2mm]node2.west);
	\draw[arrow] ([yshift=-2mm]node2.west) -- ([yshift=-2mm]node1.east);

	\draw[arrow] ([yshift=2mm]node2.east) -- node[anchor=south] {Cond.~\eqref{eqn:lin subset wig cond}} ([yshift=2mm]node3.west);
	\draw[arrow] ([yshift=-2mm]node3.west) -- ([yshift=-2mm]node2.east);

	\draw[arrow] ([yshift=2mm]node3.east) -- node[anchor=south] {Cond.~\eqref{eqn: lin subset jor cond}} ([yshift=2mm]node4.west);
	\draw[arrow] ([yshift=-2mm]node4.west) -- ([yshift=-2mm]node3.east);
	\end{tikzpicture}
\end{center}
	If there is a real structure on $\H$, and if $\Span(Q)=\Sym(Q)$, then $\JorR(Q)=\Jor(Q)$ and $\HilR(Q)=\Hil(Q)$.
\end{theorem}

In the diagram, an implication $A\Rightarrow B$ indicates that any symmetry of type $A$ is also one of type $B$.
The annotations above the three implications in the center represent conditions that have to bet met for the inclusion to hold.
The conditions \eqref{eqn:lin subset wig cond} and \eqref{eqn: lin subset jor cond} generalize the 2-design property and the 3-design property respectively.
Because they are quite technical, their detailed formulations are deferred to the proofs section.
Intuitively, the role played by the second and third moment conditions is this:
Wigner symmetries relate to inner products, which are defined by \emph{bi}linear functions;
while Jordan symmetries relate to algebraic properties, which are expressed in terms of the order-\emph{three} tensor comprising the \emph{structure constants} of the product operation.


We return to the stabilizer polytope.

Theorem~\ref{thm:main-thm} states that for the cases \ref{item:complex_jordan} and \ref{item:,d=2-rebits}, we have that
\begin{align*}
	\Kad(Q) = \Hil(Q)
\end{align*}
Combined with Thm.~\ref{thm:symmetry_summary}, this establishes that all notions of ``symmetry'' from Def.~\ref{def::symmetries} coincide for these cases.

It follows from well-known properties of stabilizer states that $0\neq \aff(Q)$ and that Condition~(\ref{eqn:lin subset wig cond}) holds for all four cases of Thm.~\ref{thm:main-thm}.
Therefore, 
\begin{align*}
	\Kad(\STP{d,n}) 
	= \Aff(\STP{d,n})
	= \Lin(\STP{d,n})
	= \Wig(\STP{d,n})
\end{align*}
in general.

It is not difficult to see that $\Kad(Q)$ is strictly larger than $\Hil(Q)$ 
if $n=1, d>2$ or 
if $d>3$.
We have therefore completed the symmetry classification.

However, there is a modified sense in which $\Kad(\STP{d,n}) = \Hil(\STP{d,n})$ does hold also in case 3, i.e.\ whenever $n>1$.
This is discussed in the next section.

\subsection{Galois-extended Clifford group}

A possible lack of elegance of the symmetry classification above lies in the fact that for $d>3, n>1$, the symplectic similitudes 
$\mr{AGSp}(\ZZ_d^{2n})\simeq\Kad(\STP{d,n})$ can no longer be implemented as Hilbert space symmetries.
However, this changes when all expansion coefficients with respect to the computational basis lie in the cyclotomic field $\QQ[\omega]$ and one allows for general Galois automorphisms, not just complex conjugation. 



The matrix elements of Clifford unitaries can be chosen to lie in $\QQ[\omega]$, so that it makes sense to consider the group generated by Clifford unitaries and the Galois automorphisms of $\QQ[\omega]$.
Such groups have been studied before, in particular, in the $n=1$ case \cite{Appleby2015GaloisUM}. 
Borrowing nomenclature from Ref.~\cite{Appleby2015GaloisUM}, we refer to them as the \emph{Galois-extended Clifford groups}, and denote them by $\Cle$.

In Sec.~\ref{sec:Galois_extended_Cl}, we will see that the adjoint action of the Galois-extended Clifford group gives rise precisely to the affine symplectic similitudes.

We then take the opportunity to look at the group structure of $\Cle$ itself, not just its adjoint representation.
An analogous problem was answered in Ref.~\cite{Appleby2015GaloisUM} for the case when $n=1$. 
Generalizing their result to $n\geq 1$, 
We note that with a suitable choice of phases, the Galois-extended Clifford group has structure
\begin{equation}
	\label{eq:group_structure_extended_Clifford_group}
	\Cle \ \simeq \
	\left( \mc{P}_{n,d} \rtimes \Sp \right) \rtimes \ZZ_d^*,
\end{equation}
where $\ZZ_d^*$ is realized by the Galois automorphisms.
In particular, the restriction to the symplectic group $\Sp$ is a true (rather than projective) representation.


%
%

We can therefore summarize our symmetry classification for the stabilizer polytope in this satisfactory way:

\begin{corollary}
	For $n>1$, the Kadison, the affine, the linear and the Wigner symmetry group of the stabilizer polytope $\STP{d,n}$ all coincide, and arise from the action of the Galois-extended Clifford group on Hilbert space.
\end{corollary}

\section{Proof of equivalences between types of symmetry}\label{sec::various_notions_of_symmetry}


\subsection{Moments of $Q$}

The discussion below will use certain \emph{moments} associated with $Q$.
These are defined with respect to a $\Kad(Q)$-invariant probability measure $\nu$ on $Q$.
In natural examples, in particular if
$Q$ is a single orbit under $\Kad(Q)$, we expect there to be a unique such measure. 
This turns out to be the case for all sets listed in Ex.~\ref{ex:Q}, though we will not make use of this uniqueness.
For the remainder of the analysis, we just assume that some $\Kad(Q)$-invariant measure $\nu$ has been chosen for each $Q$. If $\conv(Q)$ is a polytope, one can take $\nu$ to be the normalized counting measure over the vertices.
This covers all sets in Ex.~\ref{ex:Q} except for items 1 and 7.
For 1., take $\nu$ to be the one induced by the Haar measure of the unitary group, and for 6., the one induced by the Haar measure of the orthogonal group.

\begin{definition}
	With respect to the measure $\nu$ introduced above, the \emph{$k$-th moment} of $Q$ is
	\begin{align*}
		\mu_k^Q &:= \int_Q q^{\otimes k} \,\mathrm{d}\nu(q).
	\end{align*}
	Associated with the $k$-th moment is a symmetric $k$-linear form on $\Herm(\H)$:
	\begin{align}\label{eq:F^Q_k}
		F_k^Q(A_1, \dots, A_k) 
		:= \tr \big(A_1\otimes \dots \otimes A_k  \ \mu_k^Q \big)
		=
		\int_Q
		\tr \big(A_1  q\big)
		\dots
		\tr \big(A_k q\big)
		\,\mathrm{d}\nu(q).
	\end{align}
\end{definition}

The significance of the $k$-linear forms is that they are invariant under $\Lin(Q)^\dagger$:
if $\Lambda\in\Lin(Q)$, then
\begin{align*}
	F^Q_k(\Lambda^\dagger(A_1), \dots, \Lambda^\dagger(A_k)) 
	=&
	\int
	\tr \big(\Lambda^\dagger(A_1) \   q\big)
	\dots
	\tr \big(\Lambda^\dagger(A_k) \ q\big)
	\,\mathrm{d}\nu(q) \\
	=&
	\int
	\tr \big(A_1 \ \Lambda(q)\big)
	\dots
	\tr \big(A_k \ \Lambda(q)\big)
	\,\mathrm{d}\nu(q) 
	=
	F_k(A_1, \dots, A_k).
\end{align*}


To get a feeling for these moments, we will discuss them for the various examples listed in Ex.~\ref{ex:Q}.

For $Q=\mathcal{P}(\H)$, the set of all pure states equipped with Haar measure,
\begin{align}\label{eqn:k design}
	\mu^Q_k = \frac{1}{\tr P_{\Sym^k(\H)} } 
	P_{\Sym^k(\H)},
\end{align}
where
\begin{align*}
	P_{\Sym^k(\H)} = \frac1{k!} \sum_{\pi \in S_k} \pi
\end{align*}
is the projection onto $\Sym^k(\H)$, the totally symmetric subspace of $\H^{\otimes k}$.
In the formula above, $\pi\in S_k$ acts by permuting tensor factors
\begin{align*}
	\pi \left(|\psi_1\rangle \otimes \dots \otimes |\psi_k\rangle\right)
	=
	|\psi_{\pi(1)}\rangle
	\otimes \dots \otimes
	|\psi_{\pi(k)}\rangle.
\end{align*}
In general, we will say that $Q$  \emph{behaves like a complex $k$-design} if (\ref{eqn:k design}) holds.

The first moment $\mu_1^Q$
can be interpreted as the center of $Q$.
For 1-designs, we have
\begin{align}\label{eqn:1 design}
	\mu_1^Q 
	= \frac1{d} \Id,
	\qquad
	d := \dim \H
	\qquad
	\Rightarrow
	\qquad
	F_1^Q(A) = \frac1{d} \tr A.
\end{align}

For $2$-designs
we get
\begin{align}\label{eqn:f2}
	F_2^Q(A,B)
	= 
	\frac{1}{d(d+1)} 
	\big( \tr A\,\tr B + \tr A B\big) 
\end{align}
Thus, loosely speaking, $F_2^Q$ for complex 2-designs is proportional to the Hilbert-Schmidt inner product, except for ``putting a higher weight on~$\Id$''.
For $ 3 $-designs, one 
obtains (analogous to~\cite[Thm.~1]{Heinrich2019robustnessofmagic})
\begin{align*}
	F_3^Q(A,B,C) = \frac{1}{\dim(\Sym^3(\mathcal{H}))} \Big(  \tr A \tr B \tr C + \tr A \tr BC +  \tr B \tr AC +  \tr C\tr AB + \Tr(ABC)+ \Tr(ACB) \Big ). 
\end{align*}
\textbf{Example \ref{ex:Q} continued.} 
The examples 1.\ (by definition), 3., 4.\ \cite{gross-unitaries2006}, and 7.\ \cite{Sidelnikov1999} fulfill the 2-design property.
On the other hand, 2., 6.\ \cite{zhu_multiqubit_2017, webb_clifford_2016, Kueng2015} and 5.\ \cite{Hashagen2018realrandomized} and 8.\ do not.
The phase-space point operators of Ex.~8.\ form an ortho-normal basis with respect to the Hilbert-Schmidt inner product, which implies that
\begin{align*}
	F^Q_2(A_1, A_2)
	=
	\frac1{d^2} \sum_{\bold a} \tr \left( A_1  A(\bold a)  \right)\tr \left(  A(\bold a)  A_2 \right)
	=
	\frac1{d^2} \tr A_1  A_2 .
\end{align*}
In other words, $F^Q_2$ is proportional to the Hilbert-Schmidt inner product.

\subsubsection{Real spherical designs}

If $Q$ is the set of pure states that have real coefficients in some ortho-normal basis of $\H$, then $Q$ does not quite behave like a complex $k$-design.
However, the forms $F_k^Q$ can still be explicitly computed.
A somewhat tedious calculation presented in Appendix~\ref{app:spherical-designs} shows that
\begin{align}\label{eq:F^Q_2-for-real-case}
	F^Q_2(A,B)  = K \Big (\tr AB + 2 \tr A \tr B \Big )
\end{align}
and
\begin{align}\label{eq:F^Q_3-for-real-case}
	F_3^Q(A,B,C) = K_1 \tr A \tr B \tr C+K_2 \Big ( \tr A \tr BC +  \tr B \tr AC+  \tr C \tr AB \Big )+ K_3 \Big (\Tr(ABC)+ \Tr(ACB) \Big ) 
\end{align}
for positive rational numbers $K,K_1,K_2,K_3$ whose exact values are not needed here.

We say that a general set $Q\subset\Herm(\H)$ \emph{behaves like a spherical $ 4$-design}, respectively a \emph{spherical $ 6 $-design}, 
if 
Equation~(\ref{eq:F^Q_2-for-real-case})
or
Equation~(\ref{eq:F^Q_3-for-real-case})
are satisfied.

\emph{Remark:}
The form $F_k$ is thus connected to the \emph{complex $k$-design} property and the \emph{real $2k$-design} one.
The counter-intuitive factor of two between the definitions reflects an unfortunate, but established, naming convention in the literature. 

\subsection{Relations between symmetry groups}
\label{sec:symmetry-proofs}

\begin{lemma}\label{lem:jor-hil}
	$\Jor(Q)=\Hil(Q)\subset \Wig(Q)$.

	What is more, 
	if there is a real structure on $\H$, and if $\Span(Q)=\Sym(Q)$, then $\JorR(Q)=\Jor(Q)$ and $\HilR(Q)=\Hil(Q)$.
\end{lemma}

\begin{proof}
	The inclusions 
	$\Hil(Q)\subset \Jor(Q)$ 
	and
	$\Hil(Q)\subset \Wig(Q)$ 
	are trivial.

	For the converse, 
	assume $J\in \Jor(Q)$ and denote its extension to $\Herm(\H)$ by the same letter.
	Theorem~6.38 of \cite{Wan1996}
	implies
	\begin{align*}
		J(q) = \alpha U q U^\dagger
	\end{align*}
	for a non-zero multiplier $\alpha\in\RR$
	and a linear or anti-linear $U$ such that $U U^\dagger=\alpha^{-1}\Id$.
	But because $U U^\dagger$ is a positive operator, $\alpha>0$.
	Substituting $U\mapsto \sqrt\alpha U$ if necessary, we may assume that $\alpha=1$ and thus $U$ is unitary or anti-unitary.
	Hence $J\in\Hil(Q)$.

	As a linear extension of a bijection of $Q$, it holds that $J$ preserves $\Span(Q)$.
	If the latter is equal to $\Sym(Q)$, then
	an analogous argument based on Thm.~5.37 of the same reference shows that $J(q)=OqO^t$
	for an orthogonal map $O$.
\end{proof}

%
%
%




The following statement seems to be commonly known.
A related argument appears e.g.\ in the proof of Prop.~5.15 in Ref.~\cite{landsman2017}.
We re-prove it because we have not found a suitable self-contained presentation in the literature.

\begin{lemma}\label{lem:kad-aff}
  $\Kad(Q) = \Aff(Q)$.
\end{lemma}
	\begin{proof}
		Every affine map preserves convex combinations, hence $\Aff(Q)\subset\Kad(Q)$.

		Conversely, assume $K\in \Kad(Q)$ and denote its extension to $\conv(Q)$ by the same letter.
		Choose a maximal affinely independent subset 
		$S=\{q_1, \dots, q_k\}\subset Q$.
		We claim that $K$ is determined by the values it takes on $S$.
		Indeed: by maximality, any $q\in\conv(Q)$ can be expressed as an affine combination of elements from $S$.
		By collating positive and negative coefficients respectively, this decomposition can be brought into the form 
		\begin{align}\label{eqn:affine decomp}
			 q = (1+\lambda) s_1 - \lambda s_2,
			\qquad
			s_i \in \conv(S), \lambda \geq 0.
		\end{align}
		Rearranging,
		\begin{align*}
			\frac1{1+\lambda}  q
			+
			\frac\lambda{1+\lambda} s_2
			=
			s_1.
		\end{align*}
		Applying $K$ and using convexity,
		\begin{align*}
			K(s_1)
			=
			K\left(\frac1{1+\lambda}  q
			+
			\frac\lambda{1+\lambda} s_2
			\right)
			=
			\frac1{1+\lambda} K( q)
			+
			\frac\lambda{1+\lambda} 
			K(s_2)
			\quad
			\Rightarrow
			\quad
			K( q)
			=
			(1+\lambda) K(s_1)
			-\lambda K(s_2).
		\end{align*}
		Comparing with (\ref{eqn:affine decomp}) shows that $K$ commutes with affine combinations inside of $\conv(Q)$.

		Define $\Lambda$ on $\aff(Q)$ by affine extension of $\Lambda(s_i) := K(s_i)$. 
		From he above, $K$ and $\Lambda$ agree on $\conv(Q)$.
		Therefore, $K \in \Aff(Q)$.
\end{proof}


\begin{lemma}	
\label{lem:aff-lin}
	$\Lin(Q)\subset\Aff(Q)$.

	If $0\not\in\aff(Q)$, then $\Aff(Q)\subset\Lin(Q)$.
\end{lemma}

\begin{proof} 
	The first containment holds because
	every linear map is affine.

	Conversely, assume that $\Lambda\in\Aff(Q)$.
	Let 
	$S=\{q_1, \dots, q_k\}\subset Q$
	be a maximal affinely independent subset of $Q$.
	If $0\not\in\aff(Q)$, then $\{0, q_1, \dots, q_k\}$ is affinely independent in $\Span(Q)$.
	Hence $\{q_1-0, \dots, q_k-0\} = S$ is linearly independent.

	Define $L$ on $\Span(Q)$ by linear extension of $L(q_i):=\Lambda(q_i)$. 
	From the above, $\Lambda$ and $L$ agree on $\aff(Q)$.
	Therefore, $\Lambda\in\Lin(Q)$.
\end{proof}

The next claim is essentially a special case of the Mazur-Ulam Theorem.

\begin{lemma}\label{lem:lin<-wig}
	$\Wig(Q)\subset \Lin(Q)$.
\end{lemma}

\begin{proof} 
	Let $W\in\Wig(Q)$.
	Choose a maximal linearly independent set $q_1,\dots,q_k \subset Q$.
	Define $L(q_i):=W(q_i)$ and extend linearly to $\Span(Q)$.  
	Let $q\in Q$.
	We claim that $L(q)=W(q)$.
	To see this, expand $q = \sum_{j} c_j q_j$ and compute
	\begin{align*}
		\tr \big(q_i \   W(q) \big) 
		= \tr  \big(W^{-1}(q_i) \ q \big)  
		= \sum_{j} c_j \tr  \big(W^{-1}(q_i) \ q_j\big) 
		= \sum_{j} c_j \Tr \big(q_i \ W(q_j) \big) 
		= \tr  \big(q_i \ L (q)\big).
	\end{align*}
	Because the $q_i$ form a basis for the linear span of $Q$, these linear equations suffice to conclude that $L(q)=W(q)$.  
	Hence $W\in \Lin(Q)$.
\end{proof}



\subsubsection{Implications based on moments}

Let $L \in \Lin(Q)$.
It is a Wigner symmetry if it preserves the Hilbert-Schmidt form.
We already know that $L^\dagger$ preserves some symmetric linear form, namely $F^Q_2$.
Hence it is natural to suspect that the containment $\Lin(Q)\subset\Wig(Q)$ holds whenever these two forms are ``sufficiently similar''.
Inspection of Eq.~(\ref{eqn:f2}) and Eq.\eqref{eq:F^Q_2-for-real-case} show that the form $F^Q_2$ associated with a complex 2-design and real 4-design, respectively, is related to the Hilbert-Schmidt form in the sense that
\begin{align}\label{eqn:lin subset wig cond}
	\begin{split}
		&F^Q_2 \text{ is proportional to the Hilbert-Schmidt form when restricted to } \dir(Q) \\
		\text{and }\qquad&\mu^Q_1 \text{ is orthogonal to $\dir(Q)$ with respect to both $F^Q_2$ and the Hilbert-Schmidt form}
	\end{split}
\end{align}
hold.
In terms of this condition, we have the following generalization of Thm.~1~(2.) of Ref.~\cite{Heinrich2019robustnessofmagic}:

\begin{lemma}\label{lem:lin->wig}
	Condition~(\ref{eqn:lin subset wig cond}) is sufficient for $\Lin(Q)\subset \Wig(Q)$ to hold. 

	\begin{proof} 
		Assume that Eq.~(\ref{eqn:lin subset wig cond}) holds. Then we have the orthogonal decomposition
		\begin{align}\label{eq:HermH_decomposition}
			\HermH =  
			\Span(\{\muf\}) 
			\oplus\dir(Q) \oplus \Span(Q)^\perp,
		\end{align}
		where $\Span(Q)^\perp$ is the orthogonal complement to $\Span(Q)$ with respect to the 
		Hilbert-Schmidt
		inner product. 
		Let $L\in\Lin(Q)$ and denote its extension to $\Span(Q)=\Span(\{\muf\}) \oplus\dir(Q)$ by the same letter.
		Further extend $L$ to all of $\HermH$ by letting it act as the identity on $\Span(Q)^\perp$.

		Because $L$ is a bijection of $Q$, it preserves 
		$\muf$ and 
		$\dir(Q)$.
		Therefore, $L$ is block diagonal 
		\begin{align*}
			L=1 \oplus (L \restriction_{\dir(Q)})\oplus \Id
		\end{align*}
		with respect to the decomposition (\ref{eq:HermH_decomposition}).
		Because the spaces are orthogonal to each other, the adjoint $L^\dagger$ can be taken block-wise.
		But the adjoint of the restriction of $L$ to $\dir(Q)$ is an orthogonal map by the first part of (\ref{eqn:lin subset wig cond}).
		Hence $L\in\Wig(Q)$.
	\end{proof}
\end{lemma}

\emph{Remark:} 
If $\Lin(Q)$ is transitive on $Q$, and also acts irreducibly on $\dir(Q)$, then one can see that (\ref{eqn:lin subset wig cond}) is also necessary for $\Lin(Q)\subset \Wig(Q)$ to hold.  

The product operation on an algebra $\Acal$ is a linear map $\Acal\otimes\Acal\to\Acal$, i.e.\ an element of
 $\Acal^*\otimes\Acal^*\otimes\Acal$ (c.f.\ Ref.~\cite{landsberg2011tensors}).
(The coefficients of this third-order tensor with respect to a basis of $\Acal$ are known in physics as the \emph{structure constants} of $\Acal$).
If $L\in\Lin(Q)$, then $L$ preserves some third-order tensor, namely $F^Q_3$.
As above, we should therefore not be surprised to find that $\Lin(Q)\subset\Jor(Q)$ whenever $F^Q_3$ is ``sufficiently similar'' to the tensor describing the Jordan product.

Indeed, for the containment $\Lin(Q) \subset \Jor(Q)$ to be true, we give sufficient conditions by augmenting 
Eq.~(\ref{eqn:lin subset wig cond})
with constraints on $F_3^Q$ as follows:
\begin{align}\label{eqn: lin subset jor cond}
	\begin{split}
		&\text{ Eq.~(\ref{eqn:lin subset wig cond}) holds }  \\
		\text{and }\qquad& 
		\muf \propto \mathbbm{1}   \\
		\text{and }\qquad& 
		\Span(Q)=\HermH
		\text{ or }
		\Span(Q)=\Sym(\H),
		\\
		\text{and }\qquad& F^Q_3(A,B,C) \text{ is proportional to } \Tr(ABC)+\Tr(ACB) \text{ when restricted to } \dir(Q).
	\end{split}
\end{align}
One verifies that if $Q$ behaves like an complex $3$-design, or a real $6$-design, then it satisfies the above. 

We then have the following generalization of Thm.~1~(3.) of Ref.~\cite{Heinrich2019robustnessofmagic}:
\begin{lemma}\label{lem:lin->jor}
	Condition \eqref{eqn: lin subset jor cond} is sufficient for $ \Lin(Q) \subseteq \Jor(Q) $ to hold. 
\end{lemma}

\begin{proof}
	Let $L\in\Lin(Q)$ and denote its extension to $\Span(Q)$  by the same letter.

	We collect a few preliminary consequences of 
	\eqref{eqn: lin subset jor cond}, first for the case where $\Span(Q)=\Herm(\H)$:
	\begin{itemize}
		\item
			$\dir(Q)=\Herm_0(\H)$, the space of trace-free Hermitian matrices
			(by the second part of \eqref{eqn:lin subset wig cond} and the second part of \eqref{eqn: lin subset jor cond}),
	\end{itemize}
	and by inspection of the proof of Lem.~\ref{lem:lin->wig}:
	\begin{itemize}
		\item
			$L^\dagger=L^{-1}$ and thus $L^\dagger$ implements a symmetry in $\Lin(Q)$,
		\item
			therefore both
			$L$ and $L^\dagger$ preserve $\Id$, $\Herm_0(\H)$, and are orthogonal maps when restricted to $\Herm_0(\H)$.
	\end{itemize}

	Now let $A, B, C\in\Herm(\H)$.
	Write $A=A_0+a\Id$ for $A_0\in\Herm_0(\H)$ and expand $B, C$ likewise.
	Then
		\begin{align*}
	  \tr A \big( B\circ C \big)
		&=
		\frac12
		\big(
			\tr A_0B_0C_0
			+
			\tr A_0C_0B_0
		\big)
		+
		a \tr B_0C_0 +  b \tr A_0C_0 +  c \tr A_0B_0
		+
		abc \tr \Id.
		\end{align*}
		Each of the summands on the right hand side is preserved by $L$ and thus
		\begin{align*}
			\tr \big( A ( B \circ C ) \big)
			=
			\tr \big( L(A) ( L(B) \circ L(C) )\big)
			=
			\tr \big(A\,L^\dagger ( L(B) \circ L(C) )\big).
		\end{align*}
		Because $A$ is arbitrary, it follows that
		\begin{align*}
		  B \circ C
			=
			L^\dagger( L(B) \circ L(C) )
			\quad\Rightarrow\quad
			L(
			B \circ C)
			=
			L(B) \circ L(C) 
		\end{align*}
		and so $L\in\Jor(Q)$.

		The case $\Span(Q)=\Sym(\H)$ follows analogously, making use of the second part of Lem.~\ref{lem:jor-hil}.
\end{proof}

We have thus established all relations claimed to hold in Thm.~\ref{thm:symmetry_summary}:
\begin{center}
	\begin{tikzpicture}[node distance=3cm]
		\node (node0) [small] {$\Kad$};
		\node (node1) [small, right of=node0] {$\Aff$};
		\node (node2) [small, right of=node1] {$\Lin$};
		\node (node3) [small, right of=node2] {$\Wig$};
		\node (node4) [small, right of=node3] {$\Jor$};
		\node (node5) [small, right of=node4] {$\Hil$.};

		\draw[darrow] (node0.east) -- node[anchor=south] {Lem.~\ref{lem:kad-aff}} (node1.west);
		\draw[darrow] (node1.east) -- node[anchor=south] {Lem.~\ref{lem:aff-lin}} (node2.west);

		\draw[arrow] ([yshift=2mm]node2.east) -- node[anchor=south] {Lem.~\ref{lem:lin->wig}} ([yshift=2mm]node3.west);
		\draw[arrow] ([yshift=-2mm]node3.west) -- node[anchor=north] {Lem.~\ref{lem:lin<-wig}} ([yshift=-2mm]node2.east);

		\draw[arrow] ([yshift=2mm]node3.east) -- node[anchor=south] {Lems.~\ref{lem:lin<-wig}, \ref{lem:lin->jor}} ([yshift=2mm]node4.west);
		\draw[arrow] ([yshift=-2mm]node4.west) -- node[anchor=north] {Lem.~\ref{lem:jor-hil}} ([yshift=-2mm]node3.east);

		\draw[darrow] (node4.east) -- node[anchor=south] {Lem.~\ref{lem:jor-hil}} (node5.west);
	\end{tikzpicture}
\end{center}

\section{Proof of the main theorem}\label{sec:proof-main-thm}

\subsection{ $n=1$}
First, we will consider the case of one qudit and prove Statement~\ref{item:n=1} of our main theorem. 
\begin{theorem}\label{thm::SSymOne}
  Let $d$ be an odd prime. Then, the linear symmetry group of the stabilizer polytope \STP{d,1} is isomorphic to the wreath product of $S_d$ by $S_{d+1}$,
  \begin{equation}
    \LinP{\STP{d,1}}\cong S_d\wr S_{d+1}.
  \end{equation}
\end{theorem}
We will present two ways to arrive at this result. The first one makes use of the symmetry considerations in Sec.~\ref{sec::various_notions_of_symmetry}, while the second one uses the theory of polytopes. 

\subsubsection{Proof via Wigner symmetries}

For $n=1$, the set of stabilizer states can be grouped into $d+1$ mutually unbiased bases (MUBs), each of which contains $d$ states. This well-known fact is equivalent to
\begin{equation}\label{eq::ip_one}
 \tr  \left( \Pi_{L}^{g} \ \Pi_{M}^{h}\right) = \begin{cases}
  	1, &\text{if } L = M \text{ and } g = h  \\
  	0, &\text{if } L = M \text{ and } g \neq h \\
  	\frac{1}{d} &\text{if } L \neq M, 
  \end{cases}
\end{equation}
where $ L,M \subset \ZZ_{d}^2$ are (one-dimensional) Lagrangian subspace and $ g: L \to \ZZ_{d}, \, h:M \to \ZZ_{d} $ are additive functions on $ L $ respectively $ M $.  
Since $ d+1 $ mutually unbiased bases, and $ 1 $-qudit stabilizer states as a special instance of that, form a complex projective 2-design, we are free to use Wigner's definition of symmetry groups.
Recall from Def.~\ref{def::symmetries} that a Wigner symmetry is a permutation of \Stab{1,d} that preserves inner products. Equation~\eqref{eq::ip_one} immediately implies that Wigner symmetries are exactly the permutations of \Stab{1,d} that map bases to bases. Any such permutation is specified by the choice of one permutation from $S_d$ for each basis (permuting its elements) and one permutation from $S_{d+1}$ (permuting the bases). 
It follows directly that the resulting symmetry group is the wreath product $S_d \wr S_{d+1}$.

\subsubsection{Proof via direct sum decomposition}
	
For the second way to prove Thm.~\ref{thm::SSymOne}, it will be convenient to shift the stabilizer polytope from the affine subspace of trace one operators $\Herm_1(d)$ to the linear subspace of traceless operators $\Herm_0(d)$ by applying the affine map
\begin{equation*}
  \mr{M}:\Herm_1(d)\rightarrow\Herm_0(d),\,X\mapsto\mr{M}(X)=X-\frac{1}{d}\Id.
\end{equation*}
We denote the shifted polytope by $\stp{d,1}$ and use the shorthand notation $ \ssc{L}{g}=\mr{M}(\SSC{L}{g}) $. The main insight needed to prove Thm.~\ref{thm::SSymOne} is that this polytope decomposes as a direct sum. 

\begin{lemma}\label{lem::DirectSumDecomp}
  The shifted stabilizer polytope $ \stp{d,1} $ decomposes as a direct sum 
  \begin{equation}\label{eq::sum_dec_n1}
    \stp{d,1}=\bigoplus\limits_{L\in\mc{G}_{1,d}}\Delta_d(L),
  \end{equation}
  where $\Delta_d(L)=\mr{conv}(\{\pi_{L}^{g}\,:\,g\in L^\ast\})$ is the centered regular $d$-simplex associated to a stablizer basis and $\mc{G}_{1,d}$ is the set of all lines in $\ZZ_d^2$.
\end{lemma}

Note that such a statement always holds when the considered polytope is the convex hull of MUB projectors. 
The convex hull of $ d+1 $ MUB projectors is also known as the \emph{complementarity polytope}~\cite{Bengtsson2004}. 
A proof of Lem.~\ref{lem::DirectSumDecomp} can be found in \cite[Sec.~12.4]{Bengtsson2006}, however, we include it here for convenience. 
\begin{proof}
Due to Eq.~\eqref{eq::ip_one}, the inner product between two vertices of $ \stp{d,1} $ can only attain only three possible values:
\begin{equation}\begin{aligned}\label{eq::trip_ssc_n1}
    \trip{\ssc{L}{g}}{\ssc{M}{h}}=\frac{1}{d}
    \begin{cases}
      0,&\text{ if } L\neq M, \\
      -1,&\text{ if } L=M \text{ and } g\neq h, \\
      d-1,&\text{ if } L=M \text{ and } g=h.
    \end{cases}
  \end{aligned}
	\end{equation}
  For each ray $L$, the corresponding vertices sum to zero
  \begin{equation*}
    \sum\limits_{g\in L^\ast}\ssc{L}{g}=0,
  \end{equation*}
  which, when combined with Eq.~\eqref{eq::trip_ssc_n1}, implies that their convex hull 
  \begin{equation}
    \Delta_d(L):=\mr{conv}\{\ssc{L}{g}:g\in L^\ast\},
  \end{equation}
  is a centered regular $d$-simplex. What is more, the fact that all of these simplices are mutually orthogonal c.f., Eq.~\eqref{eq::trip_ssc_n1}, implies that the shifted stabilizer polytope decomposes as a direct sum
\end{proof}

We can use Lemma \ref{lem::DirectSumDecomp} to obtain a hyperplane description of $ \STP{d,1} $, as originally proven in \cite[Thm.~1]{gott06class}.
\begin{corollary}
  Let $ L_1,\ldots,L_{d+1}  $ be the set of all $ d+1 $ distinct lines in $ \ZZ_d^2 $. Then
  \begin{equation*}
    \stp{d,1}= \left\{ A \in \Herm_1(d) \,:\, \Tr\left(AX\right) \ge 0 \text{ for all } X \in \mathcal{F}_d \right\},
  \end{equation*}
  where $ \mathcal{F}_d $ is given by 
  \begin{align}\label{eq::facet-description-n=1}
    \mathcal{F}_d = \left\{  X  =\frac{1}{d} I + \sum_{i = 1}^{d+1} \pi_{L_i}^{g_i} \,:\, (g_1,\ldots,g_{d+1}) \in L_1^\ast \times \cdots \times  L_{d+1}^\ast  \right\}.
  \end{align}
\end{corollary}
\begin{proof}
  It follows from Eq.~\eqref{eq::trip_ssc_n1} that the polytope $\Delta_d(L)$ is a self-dual simplex in the linear subspace $ U_L := \text{span}_{\mathbb{R}} \{ \pi_L^g \,:\, g \in L \} \subset \Herm_0$. Therefore, it holds that
  \begin{equation*}
    \Delta_d(L) = \left\{ A \in U_L  \,:\, \Tr\left(A\pi_L^g\right) \ge -\frac{1}{d} \text{ for all } g \in L^\ast \right\}.
  \end{equation*}
  Since the facets of a direct sum of polytopes are given by all possible sums of the facets of the individual summands~\cite[Lem.~7.7]{ziegler2012lectures}, Lem.~\ref{lem::DirectSumDecomp} implies that 
  \begin{align*}
    \stp{d,1} = \left\{ A \in \Herm_0(d) \,:\, \Tr \Big (\sum_{i = 1}^{d+1} \pi_{L_i}^{g_i} A \Big ) \ge -\frac{1}{d}\text{ for all } (g_1,\ldots,g_{d+1}) \in L_1^\ast \times \cdots \times L_{d+1}^\ast\right\},
  \end{align*} 
  which directly yields a hyperplane description of $ \STP{d,1} $, as stated in the corollary. 
\end{proof}

To describe the linear automorphism group \Aut{\stp{d,1}}, we can adapt an existing result on symmetries of direct sums. This result stems from the combinatorial theory of abstract polytopes c.f., Proposition 7.2 in \cite{GLEASON2018287}. 
\begin{lemma}\label{lem::combsymmetries}
  Let $P$ be a polytope that does not admit a decomposition of the form $ P = P_1 \oplus P_2 $ for $ P_1,P_2 \neq P $ and let $Q=\bigoplus_{i=1}^{m}P$. Then, the combinatorial symmetry group of $Q$ is given by
  \begin{equation}
    \mr{Aut}(Q)\cong\times_{i=1}^m\mr{Aut}(P)\rtimes S_m =\mr{Aut}(P)\wr S_m.
  \end{equation}
\end{lemma}
In Theorem 7.9 of \cite{GLEASON2018287}, it is shown that simplices statisfy the prerequisites of Lemma \ref{lem::combsymmetries}, and it is well-known that the combinatorial symmetry group of a $d$-simplex is $S_d$, where $S_d$ denotes the symmetric group on $d$-symbols.
Due to the fact that all combinatorial symmetries of a centered $ d $-simplex can be linearly realized, Lem.~\ref{lem::DirectSumDecomp} and \ref{lem::combsymmetries} yield the following corollary.
\begin{corollary}
  The affine symmetry group of $ \stp{d,1} $ is given by 
  \begin{equation}\label{eq::stab_sym_n1_int0} 
    \Aut{\stp{d,1}}\cong S_d\wr S_{d+1}.
  \end{equation}
	In particular, since shifting does not change the geometry: 
	\begin{equation}\label{eq::stab_sym_n1_int} 
		\Aut{\STP{d,1}}\cong S_d\wr S_{d+1}.
	\end{equation}
\end{corollary}
%
%

\subsection{Rebit stabilizer states}

To prove Statement~\ref{item:,d=2-rebits} of our main theorem, it suffices to observe that real stabilizer states form a spherical $6$-design \cite{Sidelnikov1999}. 
(In fact, any set of vectors that is given as an orbit under the real Clifford group has this property \cite{nebe2006self}, see also \cite{Hashagen2018realrandomized}).
Hence, every Kadison symmetry $K$ of the set of real stabilizer states
	$\{  C\ket{0} \, : \, C \in \mathcal{C}_{2,n,\RR}  \}$
	is also a Jordan symmetry on $\Sym(\H)$, and therefore, due Lem.~\ref{lem:jor-hil}, it follows that $ K $ is also a Hilbert space symmetry of the form
\begin{align*}
	K(q) = OqO^{\sf t}
\end{align*}
for an orthogonal matrix $ O $. 
Because $\mathcal{C}_{2,n,\RR}$ is maximal in $\mr{GL}\left(2^n,\RR\right)$ (see \cite[Thm.~5.6]{Nebe2001}), it must hold that $O\in \mathcal{C}_{2,n,\RR}$.

\subsection{$d$ odd, $n>1$}
In this section we prove  Statement~\ref{item:n>1,d-odd} of our main theorem. In other words, we show that for $ n > 1 $ and $d$ odd, any stabilizer symmetry is of the form
\begin{equation}
	\begin{aligned}\label{eq::affSim_on_PSP}
		A(\bold{b})\mapsto A(R\bold{b}+\bold{a}), \quad \forall\, \bold{b}\in \ZZ_d^{2n}
	\end{aligned}
\end{equation}
for some symplectic similitude $R \in \mr{GSp}(\ZZ_d^{2n})$ and a vector $\bold{a} \in \ZZ_{d}^{2n}$. 

The proof will proceed in two steps. First, we will show that stabilizer symmetries act bijectively on phase-space point operators. Then, we show that the action is in fact realized by an affine symplectic similitude.
\begin{lemma}\label{lem::SSym_act_on_A}
  Let $\Phi\in\LinP{\STP{d,n}}$ be a linear stabilizer symmetry.  Then, there exists a bijection $\varphi:\ZZ_d^{2n}\rightarrow\ZZ_d^{2n}$ such that for any $\bold{b}\in \ZZ_d^{2n}$ it holds that
  \begin{equation}\label{eq::small-phi-acting-on-Z_d2n}
    \Phi\left[A(\bold{b})\right]=A\bigr(\varphi(\bold{b})\bigl).
  \end{equation}
\end{lemma}
\begin{proof}
According to Lemma \ref{lem:lin->wig}, any affine stabilizer symmetry preserves the trace inner product. 
To connect this to the language of symplectic geometry we can express the inner product between stabilizer states in terms of the intersection between Lagrangian subspaces and a comparison of characters
\begin{equation}\label{eq::ip_HW}
 \tr   \left( \Pi_{L}^{g} \ \Pi_{M}^{h}\right)=d^{\dim{(L\cap M)}-2n} \ \delta\left[g\vert_{L\cap M}=h\vert_{L\cap M}\right].
\end{equation}
Now it is easy to see that the restrictions of a single linear functional $\hat{f}: \ZZ_d^{2n} \to \ZZ_d$, can be used to construct a maximal set of mutually non-orthogonal stabilizer states as follows
\begin{equation}\label{eq::def_Sf}
  S_{\hat{f}}=\left\{ \Pi_L^{\hat{f}\vert_L}\,\bigg\vert\,L\subset \ZZ_d^{2n} \text{ Lagrangian}\right\}.
\end{equation}
The first step is to prove that stabilizer symmetries define a group action on the set $ \{ S_{\hat{f}} \, \vert \, \hat{f} \in (\ZZ_d^{2n})^\ast \} $. This shows that any stabilizer symmetry $\Phi$ induces a bijection on the dual of the phase space $(\ZZ_d^{2n})^\ast$, which we will turn into the desired map $\varphi$, using the symplectic form.

We begin by choosing a linear functional $\hat{f}\in (\ZZ_d^{2n})^\ast$ and a linear stabilizer symmetry $\Phi\in\LinP{\STP{d,n}}$. Since stabilizer symmetries preserve inner products, we immediately have that the states in the image of $S_{\hat{f}}$ have nonzero mutual inner products. Together with the definition of $S_{\hat{f}}$ in Eq.~\eqref{eq::def_Sf} this implies that the image also contains exactly one stabilizer state for each Lagrangian subspace
  \begin{equation*}
    \Phi\left[S_{\hat{f}}\right]=\left\{\Pi_L^{g_L}\,\big\vert\,L\subset \ZZ_d^{2n} \text{ Lagrangian}\right\}.
  \end{equation*}
A priori it is not clear that the linear functionals $g_L:L\rightarrow\ZZ_d$ are restrictions of some global functional $\hat{g}\in (\ZZ_d^{2n})^\ast$. However, according to Eq.~\eqref{eq::ip_HW}, non-zero inner products imply that
\begin{equation*}
	\begin{aligned}
		g_L\vert_{L\cap M}=g_M\vert_{L\cap M}\,,\quad\forall\,L,M\subset\ZZ_d^{2n}\text{ Lagrangian}\,.
	\end{aligned}
\end{equation*}
Thus, we have that the function $\hat{g}:\ZZ_d^{2n}\rightarrow\ZZ_d,\,\bold{a}\mapsto \hat{g}(\bold{a}):=g_L(\bold{a})$, where $\bold{a}\in L$, is well-defined. What is more, $\hat{g}$ is linear when restricted to any Lagrangian subspace of $\ZZ_d^{2n}$. For $ n > 1 $, this property is sufficient to conclude that $\hat{g}$ is in fact linear globally, c.f. Lemma 1 in \cite{Delfosse_2017}, i.e., that
\begin{equation*}
    \Phi\left[S_{\hat{f}}\right]=\left\{\Pi_L^{\hat{g}\vert_L}\,\bigg\vert\,L\subset \ZZ_d^{2n}\text{ Lagrangian}\right\}=S_{\hat{g}}.
\end{equation*}
We can now define the the bijection $\varphi:\ZZ_d^{2n}\rightarrow \ZZ_d^{2n}$ as the map on phase space that satisfies
  \begin{equation*}
    \Phi\left[S_{\left[\bold{b},\cdot\right]}\right]=S_{\left[\varphi(\bold{b}),\cdot\right]}, \quad\forall\;\bold{b}\in\ZZ_d^{2n}.
  \end{equation*}
What remains to show, is that the map $\varphi$ gives the action of the stabilizer symmetry $\Phi$ on phase space point operators. First, note that the sum over the set $S_{\hat{f}}$ from Eq.~\eqref{eq::def_Sf} is given by 
\begin{equation}\label{eq::sum_Sf}
  \sum\limits_{\Pi\in S_{\left[\bold{b},\cdot\right]}}\Pi=C\,\bigl(\Id+A(\bold{b})\bigr),
\end{equation}
where $C$ is is a strictly positive constant that is independent of $\bold{b}$. By applying Eq.~\eqref{eq::sum_Sf} we get
  \begin{equation*}
    \sum\limits_{\Pi\in S_{\left[\bold{b},\cdot\right]}}\Phi\left[\Pi\right]=\sum\limits_{\Pi\in S_{\left[\varphi(\bold{b}),\cdot\right]}}\Pi=C\,\bigl(\Id+A(\varphi(\bold{b}))\bigr).
  \end{equation*}
  Calculating the same sum again, this time using the fact that stabilizer symmetries are unital, c.f. Eq.~\eqref{eqn:1 design}, yields
  \begin{equation*}
    \sum\limits_{\Pi\in S_{\left[\bold{b},\cdot\right]}}\Phi\left[\Pi\right]=C\,\bigl(\Id+\Phi\left[A(\bold{b})\right]\bigr),
  \end{equation*}
  which concludes the proof.
\end{proof}
The second part can be split into two tasks. First, we confirm that $\varphi$, as defined in Eq.~\eqref{eq::small-phi-acting-on-Z_d2n}, is affine by showing that it permutes affine lines. (By the fundamental theorem of affine geometry \cite[Chapter~9, Thm.~15]{bennett2011affine}, these two properties are equivalent.)
\begin{lemma}
  Let $\Phi\in\LinP{\STP{d,n}}$ be a linear stabilizer symmetry. Then, the map $\varphi$ permutes affine lines.
\end{lemma}
\begin{proof}
Let $l\subset \ZZ_d^{2n}$ be a one dimensional subspace and let $\bold{a}\in \ZZ_d^{2n}$ be a vector. Then, it is always possible to construct two Lagrangians $L$ and $M$ such that their intersection is equal to $l$. Shifting them by the vector $\bold{a}$ yields affine Lagrangian subspaces that intersect in $l+v$. Therefore we have
  \begin{equation*}
    \varphi(l+\bold{a})=\varphi\bigl((L+\bold{a})\cap(M+v)\bigr)=\varphi(L+\bold{a})\cap\varphi(M+\bold{a}),
  \end{equation*}
  where the r.h.s. is an affine line since $\varphi$ permutes affine Lagrangian subspaces and the intersection of affine subspaces is again an affine subspace.
\end{proof}
The second task is to show that the linear part of $\varphi$ is a symplectic similitude. This is equivalent to showing that it preserves symplectic orthogonally.
\begin{lemma}
  Let $\Phi\in\LinP{\STP{d,n}}$ be a linear stabilizer symmetry. Then, for any pair of vectors $\bold{a},\bold{b}\in \ZZ_d^{2n}$ it holds that
  \begin{equation*}
    \left[\bold{a},\bold{b}\right]=0\Leftrightarrow\left[\varphi(\bold{a})-\varphi(0),\varphi(\bold{b})-\varphi(0)\right]=0.
  \end{equation*}
\end{lemma}
\begin{proof}
  Since $\varphi$ permutes affine Lagrangian subspaces, its linear part permutes Lagrangian subspaces. For any pair of orthogonal vectors $\bold{a},\bold{b}\in \ZZ_d^{2n}$ there exists a Lagrangian subspaces $L$ that contains both. Consequently, the images $\varphi(\bold{a})-\varphi(0)$ and $\varphi(\bold{b})-\varphi(0)$, which lie in the Lagrangian $\varphi(L)-\varphi(0)$, must be orthogonal as well.
\end{proof}
Summarizing our findings, we have shown that a generic stabilizer symmetry acts on phase space point operators through an affine symplectic similitude, i.e., the property claimed in Eq.~\eqref{eq::affSim_on_PSP}. 
This concludes our proof of Statement~\ref{item:n>1,d-odd} of our main theorem. 

\section{A further extension of the Clifford group}
\label{sec:Galois_extended_Cl}

Consider the \emph{cyclotomic field} $\QQ[\omega] \subset \CC$.
For any $\alpha\in\ZZ_d^*$, the map $\omega \mapsto \omega^\alpha$
defines a Galois automorphism $C_\alpha$ of $\QQ[\omega]$.
On Weyl-Heisenberg operators, the Galois automorphism  has the following adjiont action,
\begin{align*}
	C_\alpha T(\bold a_X, \bold a_Z) C_\alpha^{-1} = T(\bold a_X, \alpha \bold a_Z)
\end{align*}
i.e.\ it implements the similitude $K_{\alpha}$ as in Eq.~(\ref{eqn:Kalpha}).
Plugging this relation into Eq.~(\ref{eq::def_PSP}) shows that $C_\alpha$ acts on the phase space point operator basis in the same way,
\begin{align}
\label{eq::Galois_automorphism_on_phase_space_point_operators}
	C_\alpha A(\bold x) C_\alpha^{-1} = A(K_\alpha \bold x),
\end{align}
and thus also permutes stabilizer states by Eq.~(\ref{eq::SS_aff}).

Thus, for vectors and operators with expansion coefficients in this subfield at least, we have found the generalization
\begin{align*}
	\Phi_{K_\alpha}: \rho \mapsto C_\alpha \rho C_\alpha^{-1}
\end{align*}
for Eq.~(\ref{eqn:conjugate linearization}) to general similitudes.

This discussion motivates introducing the \emph{Galois-extended Clifford group}.
Such an object is well-defined only if one can write the elements of the Clifford group in terms of matrices with coefficients in $\QQ[\omega]$.
Refs.~\cite{gh17,gerardin,felipe-theta} 
give the generators of the Clifford group only in terms of algebraic expression in $\omega$.

Inspecting explicit matrix representations of the generators as given in Equations (II.4)~--~(II.6) of Ref.~\cite{felipe-theta} 
shows that the coefficients are all algebraic expressions of $d$-th roots of unity, possibly multiplied by a global factor of $\left(1/\sqrt d\right)^k$ for some $k\in\NN$.
Now recall the formula for quadratic Gauss sums:
\begin{align*}
	\sum_{m=0}^{d-1} e^{i \frac{2\pi}{d} \alpha m^2} = 
	\left\{
		\begin{array}{ll}
			\left(\frac{\alpha}{d}\right) \sqrt{d},\quad &d = 1 \mod 4, \\
			i\left(\frac{\alpha}{d}\right) \sqrt{d},\quad &d = 3 \mod 4
		\end{array}
	\right.,
	\qquad
	\text{with }
	\left(\frac \alpha d\right)\text{ the Legendre symbol}.
\end{align*}
This shows that one of $\pm \sqrt d, \pm i \sqrt d$ is indeed contained in $\QQ[\omega]$ and that the application of $C_\alpha$ changes $\sqrt d$ at most by a factor of $-1$. 

The Clifford group is a normal subgroup of the Galois-extended Clifford group. This is easily seen as follows. Let $U \in \Cl$ be an arbitrary Clifford, and let it implement the transformation as given in Eq.~\eqref{eq::CL_on_PSP}. Using Eq.~\eqref{eq::Galois_automorphism_on_phase_space_point_operators}, we see that $C_\alpha U C_\alpha^{-1}$ conjugates the phase space point operator $A(\bf{b})$ as follows. \begin{equation}\label{eq:Cl_normal} \left( C_\alpha U C_\alpha^{-1} \right) \ A(\bold x) \ \left( C_\alpha U^\dag C_\alpha^{-1} \right) \ = \ A \left( K_\alpha S K_\alpha^{-1}{\bold x} + K_{\alpha} {\bold a} \right) . \end{equation} Thus an arbitrary element of the Galois-extended Clifford group may be written as \begin{align} \label{eq:arbitrary_element_Galois_extended_Clifford} \omega^\mu T(\bold a) U_S C_\alpha, \end{align} where $T(\bold a)$ is a Weyl-Heisenberg operator, $U_S$ is a Clifford operator with the action $U_S A(\bold x ) U_S^{-1} = A( S {\bold x})$. Let's denote this by $g$. Let $h=\omega^{\nu} T(\bold b) U_R C_\beta$ be another arbitrary element of the Galois-extended Clifford group, where $R \in \Sp$. Then the group composition law is given by
\begin{equation} \label{eq:ext_Cl_composition_law}
hg \ = \ \omega^{\nu + \beta \mu - \frac{1}{2} [{\bold b},R K_\beta {\bold a}]} T \left( R K_\beta {\bold a} + {\bold b}\right) U_{R K_\beta S K_\beta^{-1}} C_{\beta\alpha}.  
\end{equation}
Here we used the fact that under conjugation by a Galois automorphism, $C_\beta U_S C_\beta^{-1} = U_{K_\beta S K_{\beta^{-1}}}$, which is easily inferred by subjecting the generators in Equations (II.4) - (II.6) of Ref.\cite{felipe-theta} to the Galois automorphism, and observing that the resultant operators are other generators of the same type.
We also used the fact that for $S \in \Sp$, $U_S$ form a faithful and not a projective representation of $\Sp$ \cite{Gerardin1977,Neuhauser2002}.
Thus the group structure of the Galois-extended Clifford group, $\Cle$ is \begin{align} \label{eq:Clext_grp_strctr} \Cle \ \simeq \ \left( \mc{P}_{n,d} \rtimes \Sp \right) \rtimes \ZZ_d^*.\end{align}
To compare, the group composition of the affine symplectic similitudes (which can be extracted from the adjoint action on phase space point operators) is given as follows. \begin{equation}\label{eq:affine_similitude_group_composition_law} 
\left({\bold b}, R, \beta \right). \left( {\bold a}, S, \alpha  \right) \ = \ \left( R K_\beta {\bold a}+ {\bold b}, R K_\beta S K_{\beta^{-1}}, \beta \alpha \right). \end{equation}
Thus the corresponding group structure of that of affine symplectic similitudes is \begin{align} \label{eq:Cle_adjoin} \mr{AG}\Sp \ = \  \ZZ_d^{2n} \rtimes \mr{G}\Sp, \end{align} where, recall that $\mr{G}\Sp$ was defined in Eq.~\eqref{eq:symplectic_similitudes}. 

To compare the above with the corresponding work in \cite{Appleby2015GaloisUM}, note that there the authors examined elements of the subgroup of the Galois-extended Clifford group, 
obtained when setting $\mu=0$ and $\mathbf{a} = \mathbf{0}$ in Eq.~\eqref{eq:arbitrary_element_Galois_extended_Clifford}.
Thus $g=U_S C_\alpha$,
and $h=U_RC_\beta$.
The resulting group composition law is then given as
(see Eq.~(30) in the pre-print for \cite{Appleby2015GaloisUM})  
\begin{align}
\label{eq:group_composition_law_purely_symplectic} 
    h g \ = \ U_{R K_\beta S K_\beta^{-1}} C_{\beta \alpha},
\end{align} which agrees with the group composition law of affine symplectic similitudes given in Eq.~\eqref{eq:affine_similitude_group_composition_law} for $\mathbf{b}=\mathbf{a}=\mathbf{0}$ (Eq.~(28) in the pre-print for \cite{Appleby2015GaloisUM}). Thus, this subgroup of the Galois-extended Clifford group is seen to be isomorphic to symplectic similitudes. Such an isomorphism doesn't hold when considering the full Galois-extended Clifford group.


\section{Acknowledgements}

We thank Felipe Montealegre-Mora and Lionel Dmello for insightful discussions and for help preparing this manuscript.
This work has been supported by Germany's Excellence Strategy –
Cluster of Excellence Matter and Light for Quantum Computing (ML4Q) EXC 2004/1
- 390534769 and the German Research Council (DFG) via contract GR4334/2-2.

The author have no conflicts to disclose.

\bibliographystyle{plain}
\bibliography{bibliography}

\appendix

\section{Spherical designs in Euclidean spaces}\label{app:spherical-designs}
%

A finite set $X \subset S^{n-1} = \{ x \in \RR^n \, : \;  x^{\sf T}x = 1\}$ on the unit sphere in $\RR^n$ is called a \emph{spherical $t$-design} if the following cubature rule holds for every polynomial $p: \RR^n \to \RR$ of degree up to $t$:
\[
\int_{S^{n-1}} p(x) \, \mathrm{d}x = \frac{1}{|X|} \sum_{x \in X} p(x),
\]
where the integration is with respect to the rationally invariant measure on the unit sphere, i.e. the Haar measure.
Equivalently, $ X $ is a spherical $ t $-design if and only if~\cite{Bannai2009}:
\begin{align}\label{eq:spherical-design-strength-test}
	\sum_{x \in X} p(x) = 0  \text{ for all } p \in \Harm_k := \Big \{  p \text{ homogeneous polynomial of deg } k \, : \, 
	\Delta p := \sum_{i=1}^n \frac{\partial^2 p}{\partial x_i^2} = 0  \Big \}, \; 1 \le k \le t.
\end{align}
For more information about spherical designs and harmonic polynomials, we refer the reader to~\cite{Bannai2009,Venkov2001a,Nebe2001}. 

If $ X $ is a spherical $ 2k $-design and $ Q = \{ xx^{\sf T} \, : \, x \in X\} $, we can compute the $ k $-linear form $ F^Q_k $, as defined~\eqref{eq:F^Q_k}, explicitly. 
Therefore, observe that in this case 
\begin{align*}
	F^Q_k(A_1, \ldots, A_k) = 
	\int_Q
	\tr \big(A_1  q\big) \cdots 
	\tr \big(A_k  q\big)
	\,\mathrm{d}\nu(q) = \frac{1}{|X|}\sum_{x \in X}
	A_1[x]
	\cdots A_k[x], 
\end{align*}
where $ A[x]:= x^{\sf T} A x $.
The polynomial 
\begin{align}\label{eq:def-p}
	p(x):= A_1[x]
	\cdots A_k[x]
\end{align}
is homogeneous and of degree $ 2k $ and can be decomposed into harmonic polynomials: 
\begin{equation}
	\label{eq:harmonic-decomposition}
	p(x) = p_k(x) + \|x\|^2 p_{k-2}(x) + \|x\|^4 p_{k-4}(x) + \cdots +
	\|x\|^k p_0(x)
\end{equation}
with $p_d \in \Harm_d$ and 
$d = 0, 2, \ldots, k$. 
If  $ X $ is a spherical $ 2k $-design (or a design of higher order),~\eqref{eq:spherical-design-strength-test} implies that 
\begin{align}\label{eq:F^Q_k-and-harmonic-decomp}
	F^Q_k(A_1, \ldots, A_k) = \frac{1}{|X|}\sum_{x \in X} p(x) = \frac{1}{|X|} \sum_{x \in X} \|x\|^k p_0(x) =  \frac{1}{|X|}\sum_{x \in X}  p_0(x) 
\end{align}
To compute the constant polynomial $ p_0 $, we apply the Laplace-operator $ \Delta $ to~\eqref{eq:def-p} and~\eqref{eq:harmonic-decomposition} (similarly to~\cite[Lem.~2.1]{Heimendahl2022}).  
Euler's formula gives for a general harmonic
polynomial $q \in \Harm_d$:
\[x
\Delta \|x\|^2 q = (4d+2n)q + \|x\|^2 \Delta q = (4d+2n)q,
\]
and inductively
\begin{equation}
	\label{eq:Euler}
	\Delta \|x\|^{2(k+1)} q = (k+1) (4k+4d+2n)\|x\|^{2k} q,
\end{equation}
see for example \cite[Lemma 3.5.3]{Simon2015a}
~\footnote{The factor $2$	in (3.5.11) is not correct in \cite{Simon2015a,}; it should be $1$.}
. 
Thus, applying the Laplace-operator $ k $-times to~\eqref{eq:harmonic-decomposition} yields the following identity:
\begin{align*}
	\Delta^k(p(x)) = \Delta^k \Big (p_k(x) + \|x\|^2 p_{k-2}(x) + \|x\|^4 p_{k-4}(x) + \cdots +
	\|x\|^k p_0(x) \Big ) = K p_0(x)
\end{align*}
for some positive rational number $ K $ and $ x \in S^{n-1} $.
On the other hand, using the product rule for the Laplace operator, we have the following identities:
Setting $ \nabla = (\frac{\partial }{\partial x_1}, \ldots, \frac{\partial }{\partial x_n} ) $,
\begin{align}
	\Delta A[x] &= 2 \Tr(A)   \\
	\nabla A[x]  &= Ax   \\
	\nabla (A[x]B[x]) &=  A[x] Bx + B[x] Ax \\
	\Delta (A[x]B[x]) &= A[x] \Delta B[x] + B[x] \Delta A[x] + 2 (\nabla A[x] )^{\sf T}\nabla B[x] = 2 (\Tr( A) B[x]+ \Tr(B) A[x]+  (A^{\sf T} B)[x]  )      \\
	\Delta^2(A[x]B[x]) &= 8 \Tr(A)\Tr(B)+ 4\Tr(A^{\sf T} B)  \label{eq:laplace-twice}.
\end{align}
If $ X $ is a spherical $ 4 $-design, then Equation~\eqref{eq:laplace-twice} together with~\eqref{eq:F^Q_k-and-harmonic-decomp} shows that 
\begin{align*}
	F_2^Q(A,B) = \frac{1}{|X|} \sum_{x \in X}A[x]B[x] = \frac{1}{|X|} \sum_{x \in X} p_{0}(x) =  \frac{1}{|X|} \sum_{x \in X} K\Delta^2(A[x]B[x]) = K^\prime \Big (\tr AB + 2 \tr A \tr B \Big ).
\end{align*}
for some $ K,K^\prime > 0 $.

To compute $ F^3_Q(A,B,C) $ for $ X $ being a spherical $ 6 $-design, we have to compute 
$ \Delta^3( A[x]B[x]C[x]) $. 
Applying the Laplace operator to $ A[x]B[x]C[x] $ yields 
\begin{align*}
	\Delta(A[x]B[x]C[x]) 
	&= A[x] \Delta (B[x]C[x]) + 2 \nabla A[x] \nabla (B[x]C[x]) + A[x]B[x] \Delta A[x]  \\
	&= A[x]B[x] \Delta C[x] + A[x]C[x] \Delta B[x] + B[x]C[x]\Delta A[x] + 2(A[x] \nabla B[x] \nabla C[x] + \nabla A[x] \nabla (B[x]C[x]))
\end{align*}
and using the above identities for the Laplace operator, this equals 
\begin{align}\label{eq:laplace-once}
	2 \Big (\Tr(C)  A[x]B[x] +   \Tr(A) C[x]B[x] + \Tr(B) A[x]C[x] \Big ) + 2\Big (  
	A[x] (BC)[x]+ (B[x] (AC)[x]+ C[x] (AB)[x] )
	\Big ).
\end{align}
Applying the Laplace operator twice to~\eqref{eq:laplace-once} gives:
\begin{align*}
	\Delta^3(A[x]B[x]C[x])= Cp_0(x)= &K_1 \Tr(A)\Tr(B)\Tr(C)   \\
	+ &K_2 \Big ( \Tr(A)\Tr(BC)   
	+ \Tr(B)\Tr(AC) + \Tr(C)\Tr(AB)   \Big ) \\
	+ &K_3 (\Tr(ABC)+ \Tr(ACB))
\end{align*}
for $ K_1,K_2,K_3 > 0 $. 
Consequently, 
\begin{align*}
	F_3^Q(A,B,C) =\frac{1}{|X|} \sum_{x \in X} A[x]B[x]C[x] =& \frac{1}{|X|} \sum_{x \in X} K\Delta^3( A[x]B[x]C[x])     \\
	=& \; \; \; \; \,K_1^\prime \Tr(A)\Tr(B)\Tr(C)   \\
	&+ K_2^\prime \Big ( \Tr(A)\Tr(BC)   
	+ \Tr(B)\Tr(AC) + \Tr(C)\Tr(AB)   \Big ) \\
	&+ K_3^\prime (\Tr(ABC)+ \Tr(ACB))
\end{align*}
for $ K, K_1^\prime,K_2^\prime,K_3^\prime > 0$.

In summary, this shows that if $ X $ is a spherical $ 4 $- or $ 6 $-design, then the set $ Q = \{xx^{\sf T} \, : \, x \in X\} $ behaves like a real $ 4 $- or $ 6 $-design in the sense of Equations~\eqref{eq:F^Q_2-for-real-case} and~\eqref{eq:F^Q_3-for-real-case}.

\section{The symmetry group for a single qubit}
\label{app:B}
By Thm.~\ref{thm:main-thm}, the symmetry group for a single qubit is special, as it is both a wreath product and realized by the adjoint action of the extended Clifford group.
As a consistency check, we directly verify that the extended Clifford group coincides with $S_2 \wr S_3$.
 This is done in Tab.~\ref{tab:d2n1}.
\begin{table}
    \centering
    \begin{tabular}{c|l}
        Extended $\Cl$ & $S_2 \wr S_3$ \\ \hline 
        Complex conjugation &~$\left[e;e^X,\tau^Y,e^Z\right]$ \\
        Conjugation by Y &~$\left[e;\tau^X,e^Y,\tau^Z\right]$ \\
        Conjugation by Z&~$\left[e; \tau^X,\tau^Y,e^Z\right]$ \\
        Conjugation by $H$&~$\left[ \tau^{XZ}; e^X, \tau^Y, e^Z \right]$ \\
        Conjugation by $S$&~$\left[ \tau^{XY}; e^X, \tau^Y, e^Z  \right]$ \\
    \end{tabular}
    \caption{Generators of the extended $\Cl$ group and $S_2 \wr S_3$: the left column depicts which operation we conjugate the Heisenberg-Weyl operators by, while the right demonstrates what element of $S_2 \wr S_3$ that translates to. An arbitrary element of $S_2 \wr S_3$ is depicted as $[\sigma; g^X,g_Y,g^Z]$, where $g^X,g^Y,g^Z$ act on the $\pm$ eigenstates of the $X$, $Y$ and $Z$ operators, while $\sigma$ permutes between the $X$,$Y$,$Z$ operators (without a sign change). $e$ denote the identity group member, and $\tau$ denotes a transposition, for instance: $\tau^X$ denotes a transposition between the $+1$ and $-1$ eigenstates of $X$, whereas $\tau^{XZ}$ denotes a transposition between the $\pm$ eigenstates of $X$ and $Z$ operators (without changing the eigenvalue).}
    \label{tab:d2n1}
\end{table} 
An analogous statement does not hold for $d >2$.

\end{document}